\documentstyle[11pt]{article}
\makeatletter
\newdimen\normalarrayskip
\newdimen\minarrayskip
\normalarrayskip\baselineskip
\minarrayskip\jot
\newif\ifold \oldtrue \def\new{\oldfalse}
\def\arraymode{\ifold\relax\else\displaystyle\fi}

\def\@arrayskip{\ifold\baselineskip\z@\lineskip\z@
  \else
  \baselineskip\minarrayskip\lineskip2\minarrayskip\fi}
\def\@arrayclassz{\ifcase \@lastchclass \@acolampacol \or
\@ampacol \or \or \or \@addamp \or
 \@acolampacol \or \@firstampfalse \@acol \fi
\edef\@preamble{\@preamble
 \ifcase \@chnum
  \hfil$\relax\arraymode\@sharp$\hfil
  \or $\relax\arraymode\@sharp$\hfil
  \or \hfil$\relax\arraymode\@sharp$\fi}}
\def\@array[#1]#2{\setbox\@arstrutbox=\hbox{\vrule
  height\arraystretch \ht\strutbox
  depth\arraystretch \dp\strutbox
  width\z@}\@mkpream{#2}\edef\@preamble{\halign \noexpand\@halignto
\bgroup \tabskip\z@ \@arstrut \@preamble \tabskip\z@ \cr}%
\let\@startpbox\@@startpbox \let\@endpbox\@@endpbox
 \if #1t\vtop \else \if#1b\vbox \else \vcenter \fi\fi
 \bgroup \let\par\relax
 \let\@sharp##\let\protect\relax
 \@arrayskip\@preamble}
\makeatother

\let\ssection=\section
\def\section{\setcounter{equation}{0}\ssection}

\def\lvm{\leavevmode\hbox to\parindent{\hfill}}
\def\req#1{(\ref{#1})}

\def\BE{\begin{equation}}  
\def\EE{\end{equation} }
\def\BA{\begin{array}}   
\def\EA{\end{array}}
\def\L{\left}
\def\R{\right}

\def\mycirc{\,\raisebox{2pt}{{${\scriptscriptstyle{\circ}}$}}\,}

\def\a{\alpha}
\def\b{\beta}
\def\g{\gamma}
\def\e{\varepsilon}
\def\l{\lambda}
\def\lst{\lambda^\ast}

\def\div{\mathop{\rm div}\nolimits}

\def\asy{antisymplectic}

\def\dl#1#2{{\partial   #1 \over\partial #2}}

\def\Dl#1{{\partial\!\!\!\!\phantom{#1}\over\partial #1}}
\def\overright{\stackrel{\leftarrow}{\partial}}
\def\Dr#1#2{\,#2{\overright\phantom{l} \over\partial #1}}

\def\d{\partial}

\def\tilde{\widetilde}
\def\bar{\overline}

\def\half{{\textstyle{1\over2}}}

\def\fourth{{\textstyle{1\over4}}}

\def\cF{{\cal F}}

\def\cH{{\cal H}}

\def\cK{{\cal K}}
\def\cL{{\cal L}}
\def\cM{{\cal M}}

\def\cO{{\cal O}}

\def\cS{{\cal S}}

\def\cV{{\cal V}}
\def\cW{{\cal W}}
\def\cX{{\cal X}}

\def\oV{{\cal H}}

\def\hV{\widehat{\cal V}}
\def\tX{\widetilde{\cal X}}

\def\sK{{\sf K}}
\def\sL{{\sf L}}
\def\Pl{{\sf Pl}}

\newcommand{\ie}{{\em i.e.\ }}
\newcommand{\tPhi}{\tilde{\Phi}}
\newcommand{\ra}{{\rightarrow}}
\newcommand{\lea}{{\leftarrow}}

\newcommand{\Lra}{{\Leftrightarrow}}

\def\Goteborg{
ITP G\"oteborg 94-31\\
January 1995 }

\def\Title{{\Large\sc Triplectic Quantization:\\[5pt]
A Geometrically Covariant Description of the\\[5pt]
$Sp(2)$-Symmetric Lagrangian Formalism }}

\def\Abstract{A geometric description is given for the $Sp(2)$ covariant
version of the field-antifield  quantization of general constrained systems
in the Lagrangian formalism.  We develop  differential geometry on manifolds
in which a basic set of coordinates (`fields') have {\it two\/} superpartners
(`antifields').  The quantization on such a triplectic manifold requires
introducing several specific differential-geometric objects, whose properties
we study. These objects are then used to impose a set of generalized
master-equations that ensure gauge-independence of the path integral.  The
theory thus quantized is shown to extend to a level-1 theory formulated on a
manifold that includes antifields to the Lagrange multipliers.  We also
observe intriguing relations between triplectic and ordinary symplectic
geometry.}

\def\HEPtitlepage{%
\begin{flushright}
\Goteborg\\
{\tt hep-th@xxx}/9502031\\
{\sf Revised version}\\
27 February 1995
\end{flushright}\par
\bigskip\par
\begin{center}
\Title\\[40pt]
{\large I.~A.~Batalin}\\[0.5ex]
{\small\sl I.~E.~Tamm Theory Division, P.~N.~Lebedev Physics Institute\\[-2pt]
Russian Academy of Sciences, 53 Leninski prosp., Moscow 117924, Russia}
\\[4ex]
{\large R.~Marnelius}\\[0.5ex]
{\small\sl Institute of Theoretical Physics,
Chalmers University, S-41296 G\"oteborg, Sweden}
\\[2ex]
and\\[2ex]
{\large A.~M.~Semikhatov}\\[0.5ex]
{\small\sl I.~E.~Tamm Theory Division,
P.~N.~Lebedev Physics Institute\\[-2 pt]
Russian Academy of Sciences, 53 Leninski prosp., Moscow 117924, Russia}
\end{center}\par
\vskip4ex\par
\centerline{\sc abstract}
\vskip1.2ex
\centerline{\parbox{.9\hsize}{\addtolength\baselineskip{-2ex}
\noindent\small
\Abstract}}
}

\begin{document}
\hfuzz=1.8pt
\raggedbottom
\thispagestyle{empty}
\addtolength{\parskip}{2pt}

\HEPtitlepage

\newpage
\setcounter{footnote}{0}
\setcounter{page}{1}
\section{Introduction}\lvm
The field-antifield formalism for quantization of general dynamical systems
subjected to constraints has been developed in~\cite{BV81} and \cite{BV83},
and has found a variety of applications to quantization problems (for a
review, see, e.g.~\cite{H}).  For most applications, it was enough to work in
the special coordinates which made the field-antifield associations explicit
-- the (anti)supersymmetric version of the Darboux coordinates.

More recently, in the application to the quantization of, probably, the most
complicated field theory known, the string field theory, a {\it covariant\/}
form of the field-antifield approach proved useful \cite{HZ,SZ93}.
`Covariant' here and henceforth refers to the manifold of fields of a given
theory.  The required generalization was constructed, in several steps, in
refs.~\cite{BT93-1,BT93-2,BT94-1}; see also \cite{S} and \cite{BT94-2} for a
review and more recent developments.  `Covariantization' of the
field--antifield formalism has led to realizing the existence of, and then
solving, another interesting problem -- that of constructing a {\it
hypergauge\/} theory and of a {\it multilevel\/} generalization of the basic
BV scheme. Both these problems were solved by appropriately generalizing the
master equation.  In order to ensure gauge invariance of  the  path integral
for the partition function, an extra measure factor was identified and its
transformation properties found.  It was also found that the extra measure
factor can be naturally included into the multilevel scheme.

On the other hand, the antibracket quantization in the Lagrangian formalism
has got a powerful generalization to an {\it Sp(2)-symmetric\/} formalism
\cite{BLT-2}
--\cite{BM}\footnote{The original motivation for it being the
introduction of ghost-anti-ghost symmetry into  the Hamiltonian
BRST-anti-BRST-formalism \cite{BLT-1,BLT-ham}, see also refs.~\cite{GH1,GH2}.
}, (see also refs.~\cite{MH}
--\cite{LT}). The $Sp(2)$-symmetric formalism, however, although applicable
to a very large  class of gauge theories, has so far been limited to the
description of the field space in  `Darboux' coordinates.  It seems to us
that the recent development in the  quantization needs has, or will soon
have, come to a point requiring the two  approaches -- quantization in
arbitrary coordinates on the field space and  the $Sp(2)$-symmetric scheme --
to merge. Besides, this would look as a natural step from the point of view
of the logical development of the field-antifield quantization scheme. This
is why we decided to give in this paper an attempt for a construction of the
$Sp(2)$-symmetric Lagrangian quantization in arbitrary coordinates (as
already clear from the above, by {\sl quantization\/} we mean constructing a
finite-dimensional analogue of the path integral).

\smallskip

Introducing the $Sp(2)$-symmetric formalism, as it was implicit already in
\cite{BLT-2,BLT-3}, means departing from the well-established facts of
(anti-)symplectic geometry.  One introduces canonical {\it triplets\/}
instead of canonical pairs.  Such triplets contain two `antifields' conjugate
to one `field', each one being conjugate with respect to one of the {\it
two\/} antibracket structures (let us note in passing that when we talk about
Darboux coordinates, these would be in such a triplectic version).  These
antibrackets are degenerate, and the essence of the {\it antitriplectic\/}
(or, for short, simply {\it triplectic\/}) geometry consists in formulating
two anti-Poisson structures -- antibrackets -- whose degeneracies are related
to each other in a certain way.

Below, we will develop a differential-geometric setting for the antibrackets
and other triplectic structures in arbitrary coordinates\footnote{ We do not
claim, however, having put the formalism on arbitrary manifolds, as the
global aspects are left beyond the scope of the present paper.  This subject
is rather special also because the field-theoretic applications of the
formalism mean that the manifolds become infinite-dimensional (see, however,
\cite{S}).  Our analysis applies to a finite-dimensional model.}.  A
characteristic difference from the \asy\ case \cite{BT93-1}--\cite{BT94-2} is
that we have not only the antibrackets, but also a pair of distinguished
vector fields, whose properties must be correlated with properties of the
antibrackets and of the `odd Laplacians'. These are used to formulate the
appropriately generalized master equations.

A special and remarkable feature is that the triplectic Lagrangian
quantization also involves a relation to {\it symplectic\/} geometry.  In
fact, the geometric data required to define  triplectic quantization on an
arbitrary triplectic manifold $\cM$ include a  symplectic\footnote{{\it
Not\/} anti-symplectic!} submanifold $\cL_1\subset\cM$ (this comes together
with a Lagrangian submanifold $\cL_0\subset\cL_1$, which is the manifold of
`classical' fields of a given theory).  The appearance of a symplectic
structure, albeit on a submanifold in $\cM$, accounts for the fact that the
gauge-fixing amounts to specifying a {\it bosonic\/} function (see
ref.~\cite{BM}). The trick of the triplectic quantization is that a {\it
Poisson\/} bracket, being an even (`bosonic') operation, cannot lead to an
odd BRST-like transformation within  the statistics assignments
characteristic to the {\it Lagrangian\/} quantization.  One therefore
`encodes' the information about the symplectic structure (the Poisson
bracket) into certain fermionic vector fields $V^a$, and then develops a
version of the {\it anti\/}canonical formalism on a bigger space.  The
antibrackets on the extended space are, in a sense, a `square root' of the
Poisson bracket associated with the symplectic structure.  This provides us
by the way with a rather  non-standard reformulation of the ordinary
symplectic geometry.  Alternatively, one can say that triplectic geometry
naturally contains symplectic `subgeometry'.

Comparing the situation with what one has in the usual antisymplectic
lagrangian quantization, one should keep in mind that there the antibracket
is non-degenerate, and therefore its maximal isotropic subspaces determine
Lagrangian submanifolds.  In the triplectic case, however, (both)
antibrackets allow a large isotropic subspace of functions that has nothing
to do with defining a Lagrangian manifold.  This degeneracy is precisely
controlled by the Poisson bracket, and it is amazing that conditions that
determine Lagrangian manifolds with respect to this Poisson structure can be
recast into an {\it anti\/}symplectic form, in terms of the two brackets that
are, in a certain sense, `ghosts' of the original Poisson bracket.

\medskip

After we have formulated the basics of triplectic differential geometry, we
proceed to the quantization method.  We will introduce {\it two\/} sets of
master-equations (each one being two-component, as is always the case on a
triplectic manifold).  One of these is the `main' master equation imposed on
that part of the full action, $W$, that contains the classical action of the
theory to be quantized, the gauge generators, information about the
(non)closedness of the gauge algebra, etc.  The other master equation is
imposed on a gauge-fixing part of the master action, which will be denoted by
$X$, that contains the gauge-fixing conditions and whose job in the path
integral is ultimately to fix the gauge in such a way as to introduce no
dependence on the specific gauge chosen.  We will explain below in more
detail what precisely is meant by gauge independence in the most general
situation.  This gauge-fixing procedure and the related quantities will often
be referred to in this context as {\it hyper\/}gauge-fixing in order to
distinguish it from the traditional procedure which does not involve
antifields.  The approach that deals with two  different master equations for
two different quantities was first formulated in refs.~\cite{BT93-1,BT93-2}
for the usual antibracket quantization, while in this paper we develop this
idea in the triplectic case, where it turns out that the two master
equations, those for $W$ and $X$, are no longer two different copies of the
same equation.

Our main concern will be the equations for the (hyper)gauge-fixing part of
the master-action, $X$, and gauge independence of the partition function.
This can be studied via `fine-tuned' changes of integration variables in the
integral, which would guarantee that the integral itself does not change,
even though an effective change of $X$ (and, in particular, of the hypergauge
conditions) has been induced.  A novel point here is that the possibility
(used extensively in the previous papers on the subject) to shift the
Lagrange multipliers in the integral will now be encoded directly in the {\it
equations\/} imposed on the master-action, and this action itself will be
allowed, in principle, to have arbitrary dependence on the variables which,
provided they enter at most linearly, become the Lagrange multipliers.  This
will give the most general, to our knowledge, formulation of gauge theories.

Recall that in ordinary gauge theories, gauge invariance is understood as
actual independence of the path integral from gauge conditions or, more
generally, from the gauge fermion.  In the geometrically covariant
formulation the gauge fermion is hidden into a set of  {\it
hyper\/}gauge-fixing functions. These functions are arranged into a solution
to the `$X$'-master-equation.  Then the gauge independence takes the form of
the independence from the natural arbitrariness inherent in this
master-equation.  This applies to the \asy\ as well as the triplectic
quantization. One may think (and in the \asy\ case, prove) that this
arbitrariness in the solution of the master equation {\it effectively\/}
results in the freedom of choosing the gauge fermion encoded in the
hypergauge-fixing functions.

The new equations imposed on the gauge-fixing part of the master-action can
be neatly summarized in (and actually derived from) a higher-{\it level\/}
master equation, the one that is defined on an {\it extended\/} space that
includes antifields to the `generalized Lagrange multipliers' $\l$.  The idea
is borrowed from refs.~\cite{BT93-1,BT93-2}, but a novel feature is that,
like the rest of the theory, the sector of the antifields to $\l$ has a
triplectic structure. However, we do not give in this paper the details of
quantization of the resulting {\it level-1\/} theory (in the nomenclature
where the original theory is level 0).  Introducing the extended sector does
anyway lead to drastic simplifications, both technically and conceptually, in
studying the generalized master equations.  In particular, this enables us to
represent the automorphisms of these equations in a closed form.  The
knowledge of these automorphisms is necessary when studying admissible
deformations of the gauge-fixing part of the master-action.

The paper is organized as follows: In section~\ref{sec:2} we introduce the
main geometrical tools to be used in the antitriplectic quantization.  In
section~\ref{sec:3} the ordinary antisymplectic quantization in its most
general formulation is reviewed. This formulation serves as a guide to our
triplectic treatment in the following section. In section~\ref{sec:4} we
consider the corresponding path integral in the triplectic case and derive
the equations which must be satisfied by the master-action and the
hypergauge-fixing action in order for the path integral to be
`BRST'-invariant. In section~\ref{sec:5} we discuss gauge invariance and show
that the partition function is independent of the choice of a solution for
the hypergauge-fixing action, and in section~\ref{sec:6} we discuss a
particular ansatz that gives rise to singular gauges.  Finally we give some
concluding remarks in section~\ref{sec:7}. In two appendices we demonstrate
how to extract a Poisson bracket from the antibrackets and derive some useful
properties of variations of the introduced geometric quantities.

\section{Differential geometry of triplectic quantization:\hfill\break Main
definitions\label{sec:2}} In the following, the commutator $[\;,\;]$ denotes
the graded commutator $[A,B]=AB-(-1)^{\e(A)\e(B)}BA$.  \subsection{Reminder
on the canonical triplectic coordinates}\lvm In the standard canonical
approach, one introduces a momentum $p_A$ to each coordinate $x^A$.  In the
anti-hamiltonian formalism of refs.~\cite{BV81,BV83}, the `momentum' has
statistics opposite to that of the `coordinate', and is called antifield
$x^\ast_A$.  In the $Sp(2)$-symmetric case, there are {\it two\/} antifields
$x^\ast_{Aa}$, $a=1,2$, associated to every
coordinate~\cite{BLT-2,BLT-3,BLT-4}.  The triplets \ $(x^A,x^\ast_{Aa})$ \
can be considered as coordinates on a (neighbourhood on) a manifold $\cM$
whose dimension is therefore a multiple of three, and which can be called
`antitriplectic'.  $\cM$ is endowed with two antibrackets:  \BE (F,G)^a=
F\frac{{\stackrel{\lea}{\d}}}{{\d x^A}}\Dl{x^\ast_{Aa}}G
-(-1)^{(\e(F)+1)(\e(G)+1)}(F\leftrightarrow G)\,,\qquad a=1,2.
\label{newantibracket}\EE The symmetry properties of each of the brackets
thus introduced, are those of the usual antibracket:  \BE (F,G)^a = - (G,F)^a
(-1)^{(\e(F)+1)(\e(G)+1)}\label{symmetryproperty}\EE

The next -- and probably the most fundamental -- object of the antisymplectic
formalism is the `odd Laplacian' $\Delta$, whose properties have been widely
discussed in the literature.  In the present, triplectic, case we have two
such operators. In the `Darboux' coordinates \ $(x^A,x^\ast_{Aa})$ \ they are
\BE\Delta^a=(-1)^{\e_{\a}}\dl{}{x^A} \dl{}{x^\ast_{Aa}}\,,\qquad a=1,2\,.
\label{newDeltaflat}\EE Each of these operators is nilpotent. In fact, they
satisfy:  \BE \Delta^{\{a}\Delta^{b\}}=0\EE where the curly brackets indicate
symmetrization in the indices $a$ and $b$.

Now we are going to demonstrate how these definitions, given so far in the
`Darboux' coordinates,  generalize to arbitrary coordinates.  We will also
introduce more objects, which will be used in our quantization procedure.

\subsection{General coordinates}\lvm Let \
$\Gamma^A,\quad\e(\Gamma^A)\equiv\e_A,\quad A=1,\ldots,3M$ \ be local
coordinates on an  antitriplectic manifold $\cM$ and $\e_A\in\{0,1\}$ their
Grassmann parities.  Let us introduce (odd) tensors $E^{AB a}$, $a=1,2$, with
the properties \BE\new\BA{rcl} \e(E^{AB a})&=&\e_A+\e_B+1\\ E^{AB
a}&=&{}-E^{BA a}(-1)^{(\e_A+1)(\e_B+1)} \EA\EE

We will also need a volume form on $\cM$, which will be represented by a
scalar density $\rho$ \cite{S}.  Having such a density $\rho$, one can define
divergence of a vector field $U=U^A\d_A$ by \BE \div U = \rho^{-1}\d_A(\rho
U^A)(-1)^{\e_A}\label{divergence}\EE where $\d_A={{\d}/\d\Gamma^A}$.
Further, with both $\rho$ and $E$ at our disposal, we can define two
fermionic operators that generalize eqs.~\req{newDeltaflat}:  \BE
\Delta^a=\half(-1)^{\e_A}\rho^{-1}\d_A\!\mycirc\!\rho E^{AB a}\d_B
\label{newDelta}\EE We require them to satisfy \BE
\Delta^{\{a}\Delta^{b\}}=0\quad\Leftrightarrow\quad [\Delta^a,\Delta^b]=0\,.
\label{DeltaDelta}\EE It then follows  that $E^{AB a}$ must satisfy the
relations \BE\new\BA{l} E^{AD\{a} \d_D E^{BC|b\}}(-1)^{(\e_A+1)(\e_C+1)} +
{\rm cycle}(A\,B\,C)=0\,.\EA\label{EEidentity}\EE and \BE
(-1)^{\e_A}\rho^{-1}\d_A\Bigl(\rho E^{AB \{a}\d_B\,
(-1)^{\e_C}\rho^{-1}\d_C\bigl(\rho E^{CD|b\}}\bigr)\Bigr)=0\label{DeltaE}
\EE

Next, we introduce  two antibrackets $(\;,\;)^a$, $a=1,2$, by the formula \BE
\Delta^a(F\,G)=(\Delta^aF)\,G + F\,\Delta^aG\,(-1)^{\e(F)} +
(F,G)^a(-1)^{\e(F)}\label{DeltaFG}\EE which shows that $\Delta^a$ fails to
differentiate the product of functions on $\cM$, and it is the corresponding
antibracket that measures the deviation of $\Delta^a$ from being a
derivation.  It follows that \BE (F,G)^a={F}{{\stackrel{\lea}{\d_A}}}E^{AB
a}{{\d_B}} {G}\label{antiBG}\EE Obviously, \BE \e((F,G)^a)=\e(F)+\e(G)+1\EE
and the symmetry property \req{symmetryproperty} is fulfilled.

It now follows from \req{antiBG} that the antibrackets differentiate the
algebra of functions under multiplication:  \BE
(F,GH)^a=(F,G)^aH+G(F,H)^a(-1)^{(\e(F)+1)\e(G)}\EE Applying
eqs.~\req{EEidentity}--\req{DeltaE} as $\Delta^{\{a}\Delta^{b\}}(FG)\equiv0$
and $\Delta^{\{a}\Delta^{b\}}(FGH)\equiv0$ we arrive at \BE
\Delta^{\{a}(F,G)^{b\}}=(\Delta^{\{a}F,G)^{b\}} +
(F,\Delta^{\{a}G)^{b\}}(-1)^{\e(F)+1}\label{Deltaantibracket}\EE and \BE
((F,G)^{\{a},H)^{b\}}(-1)^{(\e(F)+1)(\e(H)+1)} + {\rm cycle}(F,G,H) =0
\label{newJacobi}\EE respectively.  As before, the curly bracket indicates
symmetrization in the indices $a$ and $b$.  Equation~\req{newJacobi} is a
version of the Jacobi identity satisfied by the two antibrackets\footnote{
Its Poisson counterpart, which says that two brackets are {\it compatible\/},
is in the basis of constructing bi-hamiltonian integrable systems
\cite{Magri,GD} and has been studied from various points of view, see,
e.$\,$g., \cite{Sts}.  In the present situation, the compatibility condition
for the antibrackets follows from eq.~\req{DeltaDelta} imposed on the
$\Delta$-operators which play a special r\^ole in the {\it anti\/}canonical
case.}.

\subsection{Antitriplectic quantities for quantization}\lvm From the results
of \cite{BM} we expect the following properties to be necessary for
$Sp(2)$-symmetric quantization of a general gauge theory: First we need an
even number of bosonic as well as fermionic fields in the theory, which in
turn requires us to consider an antitriplectic manifold $\cM$ of dimension
$6N\,$\footnote{Which is actually related to the fact, discussed below and in
Appendix~A, that there is a hidden symplectic structure in the triplectic
quantization.}.  We set therefore $M=2N$ from now on.  Next, we need odd
(fermionic) vector fields $V^a$, $a=1,2$ on $\cM$.  These are required to
satisfy \BE [\Delta^a,\,V^b]\equiv\Delta^a V^b+V^b\Delta^a=0
\label{DeltaV}\EE Applying this identity to a product of two functions $FG$
and making use of \req{DeltaFG}, we see that each $V^a$ differentiates {\it
both\/} antibrackets:  \BE V^{a}(F,G)^{b}=(V^{a}F,G)^{b} +
(F,V^{a}G)^{b}(-1)^{\e(F)+1}\label{Vantibracket}\EE Introducing the
components as \BE V^a\equiv(-1)^{\e_A} V^{A a}\d_A\,,\qquad \e(V^{A a})=\e_A
+ 1\label{Vcomponents}\EE we rewrite the conditions implied by \req{DeltaV}
as the following equations (the first of which has the usual Lie-derivative
form):  \begin{eqnarray} &&(-1)^{\e_C} V^{C a}\d_C E^{AB b}=(-1)^{\e_A}V^{A
a}{\stackrel{\lea}{\d}}_C E^{CB b} - (-1)^{(\e_A+1)(\e_B+1)} (-1)^{\e_B}V^{B
a}{\stackrel{\lea}{\d}}_C E^{CA b}\label{VEformula}\\
&&(-1)^{\e_A}\rho^{-1}\d_A\Bigl(\rho E^{AB a}\d_B V^{C b}\Bigr) +(-1)^{\e_A +
\e_B} V^{B a}\d_B\Bigl( \rho^{-1}\d_A\bigl(\rho E^{AC
b}\bigr)\Bigr)(-1)^{\e_C} =0\,.\label{VEformula2} \end{eqnarray}
Furthermore, we need  the $V^a$ to satisfy:  \BE [V^a,\,V^b]\equiv
V^{\{a}V^{b\}}=0\label{VV}\EE which is simply \BE (-1)^{\e_A}V^{A \{a}\d_A
V^{B |b\}}=0\,.\label{VVcoord}\EE

In addition to the above, we will choose $V^a$ to be divergence-free:
\BE\div V^a=0\label{divergencefree}\EE (in the sense of
eq.~\req{divergence}) in order to simplify the formulae.  This condition is
not of the same status as the previous ones however, in that it can be
removed at the expense of altering several formulae (see
Remark~\ref{divergence5} on page \pageref{divergence5}).

It will be useful to introduce the operator \BE\sK=\epsilon_{ab}V^aV^b
\label{cK}\EE that satisfies the projection property, \BE V^a\sK=\sK V^a = 0
\label{projection}\EE due to \req{VV}. In particular, \BE\sK^2=0\,.\EE

\medskip

Note that  the  properties \req{DeltaDelta}, \req{DeltaFG} and
\req{Deltaantibracket} of $\Delta^a$ are inherited by the operators \BE
\Delta_\pm^a=\Delta^a\pm{i\over\hbar}V^a\,.\label{Deltapm}\EE due to the
properties \req{DeltaV}, \req{Vantibracket} and \req{VV} of $V^a$.

\smallskip

The presence of the $V^a$ vector fields will be noticeable in many places of
our formalism, and their origin can be traced to the fact that, in triplectic
geometry, the generalized BRST-type transformations, besides being
`duplicated', acquire a `transport' term in addition to the antibracket term:
we will thus widely use transformations of the general form \BE
\delta\Gamma^A=(\Gamma^A,\cH)^a\mu_a - {\rm const}\,V^{A a}\mu_a
\label{generalizedcanonical}\EE where $\mu_a$ are odd (fermionic) parameters
or functions. The actual value of the constant is related to the
normalization of $V^a$. Note that all the defining relations for $V^a$ are
homogeneous in $V^a$. As we will see, fixing the normalization of $V^a$ has
to do with boundary conditions for generalized master equations and with the
symplectic `subgeometry' that can be extracted from the above axioms.  In
particular, although the normalization of $V^a$ is conventional, they cannot
be scaled to zero as that would mean degeneration of the Poisson structure.

\bigskip

In \cite{BM}, an $Sp(2)$-symmetric quantization was performed in `Darboux'
coordinates \BE\new\BA{l} \Gamma^A=\{\Phi^\a,\Phi^\ast_{\a a},
\bar\Phi_\a,\pi^{\a a}\}\,,\\ \e(\Phi^\a)=\e(\bar\Phi_\a)=
\e_\a\,,\quad\e(\Phi^\ast_{\a a})=\e(\pi^{\a a})=\e_\a + 1\,,
\label{DC}\EA\EE with $\Phi^\a$ being the original fields. In \cite{BM},
$\Phi^\ast_{\a a},\,\bar\Phi_\a$ were considered to be antifields while
$\pi^{\a a}$ were considered to be auxiliary field variables.  This is in
distinction to the present formulation where $\Phi^\a,\,\bar\Phi_\a$ are
considered to be field variables and $\Phi^\ast_{\a
a},\,\pi^\a_a\equiv\epsilon_{ab}\pi^{\a b}$ antifields. In our conventions
the $\Delta^a$ \req{newDelta} and the antibrackets \req{antiBG} have the
following form in terms of the Darboux coordinates \begin{eqnarray}
\Delta^a&=&(-1)^{\e_\a}\Dl{\Phi^\a}\Dl{\Phi^\ast_{\a a}}+
(-1)^{\e_\a}\Dl{\bar\Phi_\a}\Dl{\pi^{\a}_a}\label{Deltaflat}
\label{DeltaDarboux}\\ (F,G)^a&=&\biggl(F\frac{{\stackrel{\lea}{\d}}}{{\d
\Phi^\a}} \dl{}{\Phi^\ast_{\a a}}G +
F\frac{{\stackrel{\lea}{\d}}}{{\d\bar\Phi^\a}}\dl{}{\pi^{\a}_a}G-
(-1)^{(\e(F)+1)(\e(G)+1)}(F\leftrightarrow G)\biggr)\label{flatantibracket}
\end{eqnarray} in agreement with the expressions \req{newDeltaflat} and
\req{newantibracket}. The vector fields $V^a$ \req{Vcomponents} are here
given by \BE V^a=\half\epsilon^{ab}\Bigl(\Phi^\ast_{\a b}\Dl{\bar\Phi_\a} -
(-1)^{\e_\a}\pi^{\a}_b\Dl{\Phi^\a}\Bigr)\label{Vflat}\EE which is easily seen
to satisfy \req{DeltaV}.

In the present paper, however, we do not resort to the Darboux coordinates,
nor even to their existence (which is not known to us as a theorem on the
canonical form of two appropriately degenerate antibrackets and the
corresponding $V^a$ vector fields).

\subsection{A Poisson structure from triplectic geometry\label{sec:2.4}}\lvm
We are going to demonstrate the relation of the triplectic geometry to
symplectic (sub)geometry. As a motivation, note that the structure of the
field space in the Darboux coordinates \req{DC} suggests that it may be
related to a canonical phase space structure.  Indeed, for $F$ and $G$ being
two arbitrary functions of $\Phi,\bar{\Phi}$ only, we find \BE (F,
V^aG)^b=\half\epsilon^{ab}\Bigl( F\frac{{\stackrel{\lea}{\d}}}{{\d
\Phi^\a}}\dl{}{\bar\Phi_\a}G-(-1)^{\e(F)\e(G)} (F\leftrightarrow G) \Bigr)\,.
\label{Poissonbracket}\EE

It is remarkable that a Poisson structure can in fact be extracted, under a
few additional assumptions, from the general axioms introduced in
sections~2.2 and 2.3. The additional assumptions are a geometrical
counterpart of the boundary conditions that will be considered in
section~\ref{sec:4}.  They specify, among other things, non-degeneracy
properties of the tensors $E^{ABa}$ and impose certain `boundary conditions'
on the structure of triplectic manifold $\cM$.

Namely, in the triplectic manifold $\cM$, with \BE{\rm dim}\cM =
(4N-2n|2N+2n)\,,\label{Mdimension}\EE there should exist a {\it
submanifold\/} $\cL_1\subset\cM$ of dimension \BE {\rm
dim}\cL_1=(2n|2N-2n)\label{Ldimension}\EE satisfying the four requirements
listed and discussed below:\nopagebreak

\noindent (i)~non-degeneracy property:  $E^{ABa}$ should define a
non-degenerate pairing between the cotangent bundle to $\cL_1$ and conormal
bundle of $\cL_1$ in the following sense: for any
$\omega\!\in\!N\cL_1\!\subset \!T^*(\cM)$ being a 1-form that annihilates all
the vectors tangent to $\cL_1$ (i.e., $\omega(v)=0$ for $v\!\in\!T\cL_1$), we
should have \BE (\omega_AE^{ABa}\varpi_B=0\ {\rm for}\ a=1,2\ {\rm and}\
\forall\varpi\in T^*\cL_1) \Rightarrow \omega_A=0\,;\label{nondegeneracy}
\EE (ii) all functions on $\cL_1$ are annihilated by both operators
$\Delta^a$:  \BE\Delta^a F=0\,,\qquad a=1,2,\quad F\in\cF(\cL_1)\,.
\label{Deltazero}\EE Since functions on a submanifold are closed under
multiplication, we can apply $\Delta^a$ to a product of two such functions
and use \req{DeltaFG} to arrive at \BE (F,G)^a=0\,,\qquad a=1,2,\quad
F,G\in\cF(\cL_1)\label{zeroantibracket}\EE (iii) In addition, $\cL_1$ should
be such that \BE(F,V^{\{a}G)^{b\}}=0,\quad
F,G\in\cF(\cL_1)\label{iii}\EE (or equivalently \
$\Delta^{\{a}(FV^{b\}}G)=0$).  When this is satisfied, $(F,V^aG)^b$ is
$\epsilon^{ab}$ times what will become the Poisson bracket.  We thus define
\BE\{F,G\}\equiv\epsilon_{ba}(F, V^{a}G)^{b}=
\epsilon_{ab}\Delta^a(FV^{b}G)(-1)^{\e(F)}\label{PB}\EE (the last equality
follows from \req{DeltaFG}, \req{DeltaV} and \req{Deltazero}. This has then
the symmetry properties of a Poisson bracket,
$\{F,G\}=-(-1)^{\e(F)\e(G)}\{G,F\}$ and satisfies the Leibnitz rule (see
Appendix~A). Before discussing the fourth requirement that would lead
eventually to the Jacoby identity, consider in more detail the properties we
have imposed so far\footnote{Note that a related construction of extending a
given Poisson bracket to a (single!) antibracket on a bigger manifold has
been carried out in \cite{KN,A} for a particular realization of the vector
field $V$ in the `Darboux' coordinates. In the present paper, conversely, we
start with a ({\it pair\/} of) general $V^a$ vector fields and impose
conditions on $V^a$ as well as the two antibrackets that would guarantee the
existence of a Poisson subgeometry.}.

Let us specify locally the submanifold $\cL_1$ by a set of $4N$ equations
$\varphi^\mu\!=\!0$ and choose local coordinates $x^i$ on $\cL_1$.  Then
$\Gamma^A\!=\!(x^i,\varphi^\mu)$ will be local coordinates on $\cM$.  From
\req{zeroantibracket} it follows that, in such an {\it adapted\/} coordinate
system, we should have \BE E^{ija}=0\,,\qquad a=1,2,\quad i,j=1,\ldots,2N\,,
\label{zeroblock}\EE while the non-degeneracy condition \req{nondegeneracy}
takes the form \BE \omega_\mu E^{\mu ja}=0\Rightarrow\omega_\mu=0\,.
\label{minor}\EE As to \req{iii}, we write
$V^a=(-1)^{\e_i}V^{ia}\d_i+(-1)^{\e_\mu}V^{\mu a}\d_\mu$ and then \req{iii}
becomes \BE E^{i\mu a}\d_\mu V^{jb}+(a\leftrightarrow b)=0\,.\label{EV}\EE

Let us see how these conditions behave under infinitesimal coordinate
transformations \BE \delta\Gamma^A=T^A(\Gamma)=(T^i,T^\mu)\quad {\rm
with}\quad \d_\mu T^i=0\EE that consist of an arbitrary change of
coordinates on $\cL_1$ (described by $T^i(x)$) and an arbitrary
transformations of the remaining coordinates given by $T^\mu(x,\varphi)$.
First, the condition $E^{ija}\!=\!0$ varies according to \req{LieE},
resulting in \BE 0=-T^i\!\stackrel{\lea}{\d}_C E^{Cja} -E^{iCa}\d_C T^j=
-T^i\!\stackrel{\lea}{\d}_\mu\!E^{\mu ja} -E^{i\mu a}\d_\mu T^j\EE which is
satisfied by virtue of $\d_\mu T^j=0$.  Next, evaluating the variation of
\req{EV} due to \req{Tbracket} and \req{LieV}, we see that it vanishes again
by virtue of $\d_\mu T^i\!=\!0$.

In the coordinates introduced above, the candidate Poisson structure on
$\cL_1$ is determined by the tensor \BE
\omega^{ij}=\half\epsilon_{ab}(-1)^{\e_j}E^{i\mu b}\d_\mu V^{ja}\,.
\label{omegaij}\EE Requiring it to be non-degenerate would impose (in
combination with \req{nondegeneracy}) the appropriate rank conditions on the
matrix $\d_\mu V^{ja}$.

We have finally to ensure that the operation $\{F,G\}$ leaves us in the same
class of functions, namely functions on $\cL_1$.  For any two such functions
$F$ and $G$, consider $\Delta^{\{a|}(F,V^bG)^{|c\}}$, where we can use
\req{Deltaantibracket}. This vanishes due to \req{DeltaV}, hence \
$\Delta^a\{F,\,G\}=0$\@.  However, this does not automatically imply
$(H,\,\{F,\,G\})^a=0$ for $H\in\cF(\cL_1)$. This latter condition is
tantamount to \BE \epsilon_{ab}E^{k\nu c}\d_\nu(E^{i\mu b}\d_\mu V^{ja})=0\,.
\label{EEV}\EE Rewriting it (making use of \req{EV}) as \BE E^{k\nu
c}\d_\nu(E^{i\mu b}\d_\mu V^{ja})+(b\leftrightarrow c)=0\,,\EE or, in terms
of three arbitrary functions on $\cL_1$, \BE (F,(G,V^aH)^{\{b})^{c\}}=0\,,
\label{three}\EE we can show that the Jacobi identity follows for the
Poisson bracket, see Appendix~A\@.  Now, in view of the
condition~\req{nondegeneracy}, eq.~\req{EEV} states simply that
$\d_\mu\omega^{ij}=0$, which obviously guarantees that the Poisson bracket of
functions on $\cL_1$ does not acquire any dependence on the coordinates
transversal to $\cL_1$\@. It is possible, however, to impose this only as
a\nopagebreak

\noindent (iv) `boundary condition' \BE \d_\mu\omega^{ij}\Bigm|_{\cL_1}\equiv
\half\epsilon_{ab}(-1)^{\e_j}\d_\mu\bigl(E^{i\mu b}\d_\mu V^{ja}\bigr)
\Bigm|_{\cL_1}=0\label{restricted}\EE and define the Poisson bracket
accordingly, by explicitly restricting to $\cL_1$:  \BE
\{F,\,G\}=F\!\stackrel{\lea}{\d}_i\omega^{ij}\d_jG\Bigm|_{\cL_1}. \EE
Eq.~\req{restricted} will then be sufficient to guarantee the Jacobi
identity.  Thus the fourth requirement takes the form of a `boundary
condition', stating that $\epsilon_{ab}E^{i\mu b}\d_\mu V^{ja}$ has to be
constant along the directions transversal to $\cL_1$ only on the first
infinitesimal neighbourhood of $\cL_1$. In fact, the relations
\req{zeroblock}, \req{minor} (and the non-degeneracy conditions
\req{nondegeneracy}!), too, can be relaxed to those holding only on
infinitesimal neighbourhoods of $\cL_1$.

This concludes those requirements on the triplectic quantities that are
related to a fixed submanifold $\cL_1\subset\cM$\@. In the quantization of
gauge theories, this manifold serves also to impose boundary conditions on
the master actions, as will be shown below.

\section{General antisymplectic Lagrangian quantization\label{sec:3}}\lvm
Before we tackle the construction of general  hypergauge-fixed actions in the
$Sp(2)$-symmetric case using the general triplectic formulation given in the
previous section, we shall describe the corresponding construction in the
\asy\ case. In fact, even the ordinary BV theory can be given a more general
and streamlined formulation.  The \asy\ construction is presented in this
section in such a way as to make it easier to compare with the triplectic
case.  The presentation given here of the \asy/antibracket quantization
method is also (to our knowledge) the most general formulation. The reader
who is interested only in the triplectic case may proceed directly to section
\ref{sec:4}.

\subsection{The path integral and master equations}\lvm In the general
antisymplectic theory one first defines an invariant quantum master action
$W(\Gamma,\hbar)$ on an antisymplectic manifold $\cM$ by the condition that
it must satisfy the quantum master equation \BE \Delta
\exp\L(\frac{i}{\hbar}W\R)=0\;\;\Lra\;\;\;\half(W, W)=i\hbar\Delta W
\label{Wmaster}\EE where $\Delta$ and $(\;,\;)$ are the usual antisymplectic
differential and the antibracket respectively (see ref.~\cite{BT94-2}; these
structures can be thought of as being given by \req{newDelta} and
\req{antiBG} with the index $a$ dropped; however, in contrast to the
triplectic case, the antibracket must be non-degenerate).  $W(\Gamma;\hbar)$
is assumed to be expandable in $\hbar$. One then imposes the boundary
condition \BE W(\,\cdot\,;0)\Bigm|_{\cL_0}=\cS(\,\cdot\,)\label{BC}\EE where
$\cL_0$ is a Lagrangian submanifold in $\cM$ on which the action $\cS$ is
defined.  This manifold must be specified as a part of the definition of the
theory.  Further, to formulate the remaining part of the boundary conditions
on $W$, let $W^{(1)}$ be the restriction of $W(\,\cdot\,;0)$ to the first
infinitesimal neighbourhood of $\cL_0$. It is then required that \BE
(\cS,\,W^{(1)})=0\,.\label{BC1}\EE The rest of the dependence of $W$ on the
fields off the `classical' manifold $\cL_0$ is determined by the master
equation, whose solutions `propagate' from the boundary conditions \req{BC}
and \req{BC1}.

We set $\dim\cL_0=(n|N-n)$ in the following.

Next, we introduce the gauge-fixing master-action $X$ and consider the
following ansatz for the partition function path integral \BE
Z=\int\exp\biggl\{{i\over\hbar}\Bigl[W + X\Bigr]
\biggr\}\rho(\Gamma)[d\Gamma][d\lambda]\,.\label{pathintegral}\EE
$X(\Gamma,\l;\hbar)$ is assumed to be expandable both in $\hbar$ and in the
parametric variables $\l^{\a}$. The new variables $\l^{\a}$, $\a=1,\ldots,N$,
become Lagrange multipliers when the action is restricted to depend on them
linearly.  There must be $n$ bosons and $N-n$ fermions among the $\l^{\a}$ if
$\dim\cL_0=(n|N-n)$.

Now we come to imposing an equation on $X$; this will be a generalization of
the master-equation of the form \req{Wmaster}.  The generalization is due to
the fact that $X$, unlike $W$, is allowed to depend on $\l^\a$.  The sought
equation can be arrived at by requiring that the integral \req{pathintegral}
be invariant under anticanonical transformations given by \BE
\delta\Gamma^A=(\Gamma^A, -W+X)\mu\label{Btransform}\EE where $\mu$ is a
fermionic constant.  The form of \req{Btransform} is such that the resulting
conditions on $X$ will not involve $W$ when one makes use of the fact that
$W$ satisfies the quantum master equation \req{Wmaster}.  In order to obtain
as general conditions on $X$ as possible we allow, along with
\req{Btransform}, a variation of $\l$ of the form (the factor $-2$ is chosen
for convenience) \BE \delta\l^{\a}=-2R^{\a}\mu\label{ltransform}\EE where
$R^{\a}(\Gamma,\l;\hbar)$ is an a priory arbitrary function expandable in
both $\l^{\a}$ and $\hbar$, with $\e(R^\a)=\e_\a + 1$ where
$\e_{\a}\equiv\e(\l^{\a})$.  Our claim now is that invariance of the integral
\req{pathintegral} under the transformations \req{Btransform} and
\req{ltransform} will follow once we impose the equation \BE \half(X,
X)-i\hbar\Delta X - X{\!\stackrel{\lea}{\d}}_{\a}R^{\a} + i\hbar
R^{\a}{\!\stackrel{\lea}{\d}}_{\a} = 0\label{Xeq}\EE or equivalently
\BE\Bigl(\Delta-\frac{i}{\hbar}(-1)^{\e_{\a}}R^{\a}\d_{\a}-\frac{i}{\hbar}
(-1)^{\e_{\a}}\d_{\a}R^{\a}\Bigr)\exp{\L(\frac{i}{\hbar}X\R)}=0
\label{Xeqexp}\EE where $\d_{\a}\equiv\d/\d\l^{\a}$.  This is a `weak' form
of the quantum master equation, where by {\sl weak\/} we mean that the
equation contains terms other than those with the $\Delta$ operator and the
antibracket.  The name actually refers to the fact that, in the most common
case when $X$ depends on $\l^\a$ linearly via $G_\a\l^\a$, the integral is
concentrated on the hypergauge locus $G_\a=0$ and the term
$X{\!\stackrel{\lea}{\d}}_{\a}R^{\a}$ in \req{Xeq} is in fact proportional to
the $G_\a$.

Eq.~\req{Xeq} is an equation for both $X$ and $R^{\a}$. It is solved by
Taylor expanding in $\hbar$ and $\l^{\a}$.  If we first expand $X$ and
$R^{\a}$ in $\hbar$, \ie\ \BE X=X_0+i\hbar
X_1+(i\hbar)^2X_2+\ldots,\;\;\;R^{\a}=R^{\a}_0+i\hbar
R^{\a}_1+(i\hbar)^2R^{\a}_2+\ldots\label{expansion}\EE then eq.~\req{Xeq}
leads to \BE \half(X_0, X_0)=X_0{\!\stackrel{\lea}{\d}}_{\a}R^{\a}_0
\label{classXR}\EE \BE (X_0, X_1)-\Delta
X_0=X_0{\!\stackrel{\lea}{\d}}_{\a}R^{\a}_1+
X_1{\!\stackrel{\lea}{\d}}_{\a}R^{\a}_0- R^{\a}_0{\!\stackrel{\lea}{\d}}_{\a}
\label{firstXR}\EE etc. An expansion in $\l^{\a}$ leads then to a further
proliferation of equations since each power of $\l^{\a}$ must agree.

Note that by applying to the LHS of \req{Xeq} the operator \BE \Delta +
{i\over\hbar}(X,\,\cdot\,)-{i\over\hbar}R^\a(-1)^{\e_\a}\Dl{\l^\a}{}\EE we
arrive at a consistency condition \BE \Dl{\l^\a}{} \biggl(\Bigl(i\hbar\Delta
R^\a + R^\b(-1)^{\e_\b}\dl{R^\a}{\l^\b} -(X,R^\a)\Bigr)
e^{{i\over\hbar}X}\biggr)=0\label{consistency}\EE It should be realized that
this is only a member in the chain of `higher consistency relations'.  We
will now introduce a more powerful formalism that collects the different
terms in the `weak' master equation as well as {\it all\/} the higher
consistency relations into a `strong' master equation which is obtained by
introducing antifields to  $\l^\a$, as it will be done in
subsection~\ref{subsection:3.2}.

\smallskip

So far we have shown that the integral \req{pathintegral} is invariant under
the transformations \req{Btransform} and \req{ltransform} provided $X$
satisfies \req{Xeq} for appropriately chosen $R^{\a}$ in \req{ltransform}.
This is a global invariance since $\mu$ was a constant in \req{Btransform}
and \req{ltransform}.  The trick used in the BV scheme in order to prove
independence from the choice of the gauge condition consists in performing a
{\it field-dependent\/} transformation of the integration variables such that
would result in an arbitrary infinitesimal shift of the gauge conditions
within the allowed class.  Consider therefore the same transformations as
above but now with $\mu$ replaced by an arbitrary function,
$\mu\,\ra\,\mu(\Gamma,\l;\hbar)$.  Note that we allow $\mu$ to have an a
priori arbitrary dependence on $\l$. The resulting transformation of the
integral gives rise to a nonvanishing Jacobian $1+\delta J$:  \BE \delta
J=(-W+X, \mu) -2R^{\a}({\d}_{\a}\mu)(-1)^{\e_{\a}}\label{J1}\EE At this point
we need to perform an additional canonical transformation, which is obviously
non-trivial only for non-constant $\mu$:  \BE
\delta_1\Gamma^A=\frac{\hbar}{i}(\Gamma^A, \mu)\label{delta1Gamma}\EE This
has an interesting effect: it cancels the $(-W,\mu)$ term in \req{J1} while
doubling the contribution $(X,\mu)$ \footnote{From this moment on, the
$W$-part of the master action is no longer involved in the derivation. Such a
decoupling of $W$ and $X$ is the result of the opposite signs chosen in
\req{pathintegral} and \req{J1}. In other words, these two equations involve
the different operators $\Delta_\pm$ \req{Deltapm}.}.  The additional
transformation also contributes a Jacobian \BE\delta_1J=2{\hbar\over
i}\Delta\mu\label{delta1J}\EE The result of putting all terms together
amounts to the following effective change of the action in
\req{pathintegral}:  \BE \delta X= (X, f)-i\hbar\Delta
f-(-1)^{\e_{\a}}R^{\a}{\d}_{\a} f\label{deltaX}\EE where we have defined \BE
f\equiv \frac{2\hbar}{i}\mu\EE Therefore, by a transformation of dummy
variables in the integral \req{pathintegral} we have rewritten it as a
similar integral with a deformed gauge-fixing master-action $X$.

There are two basic (related) requirements to be satisfied by the new,
deformed $X$. First, it must satisfy the (weak) master equation.  One can
check explicitly that eq.~\req{Xeq} is indeed invariant under \req{deltaX}
and a simultaneous infinitesimal transformation of $R^\a$ of the form \BE
\delta R^\a=(R^\a, f) - (-1)^{\e_\b}F^{\b\a}\d_\b f\label{deltaR}\EE where
$F^{\a\b}$ appears in the solution \req{conditionfromstar} of the condition
\req{consistency}.  We will call a deformation of $X$ that goes through the
master equation {\sl consistent\/}.  Thus \req{deltaX} is a consistent
variation of $X$, in the sense that it preserves the equation imposed on $X$.

The second requirement on the variation $\delta X$ is that it describe the
{\it maximal\/} arbitrariness allowed by the equation and the boundary
conditions. This arbitrariness is then the manifestation of {\it gauge\/}
freedom inherent in our formalism.

There exists a general way to prove the desired properties of the above
variation using the $\l^*$-extended formalism.

\subsection{$\l^*$-extended sector and strong master equation
\label{subsection:3.2}}\lvm Considerable simplifications will follow when we
identify the different terms in eq.~\req{Xeqexp} as coming from a `strong'
master equation (i.e. an equation of the type of \req{Wmaster}) on a bigger
manifold.  In order to recast the weak  master equation \req{Xeqexp} into a
strong one on an extended antisymplectic manifold we introduce the
anticanonical pairs ($\l^{\a},\;\l^*_{\a}$) where  $\l^*_{\a}$
($\e(\l^*_{\a})=\e_{\a}+1$) are antifields to the variables  $\l^{\a}$. On
this extended manifold $\tilde\cM$ we may define an odd operator \BE
\Delta_{\rm ext}\equiv\Delta+(-1)^{\e_\a}\Dl{\l^\a}\Dl{\l^\ast_{\a }}
\label{Deltaextended}\EE and the corresponding antibracket \BE (F,G)_{\rm
ext}\equiv(F,G)+ \Dr{\l^\a}{F}\Dl{\lst_{\a }}{G} - \Dr{\lst_{\a
}}{F}\Dl{\l^\a}{G}\,.\label{bracketextended}\EE We define now the extended
quantum master equation by \BE \Delta_{\rm
ext}\exp{\L(\frac{i}{\hbar}\cX\R)}=0\,.\label{strongmaster}\EE The weak
master-equation \req{Xeqexp} for $X$ will now follow from \req{strongmaster}
once we expand $\cX$ in the antifields:  \BE \cX\equiv
X-\l^*_{\a}R^{\a}+\half\lst_\a\lst_\b F^{\a\b}+\cO((\l^*)^3)\EE (where
$F^{\b\a}=(-1)^{(\e_{\a}+1)(\e_{\b}+1)}F^{\a\b}$). Further, one can observe
that in the first order in $\lst$ eq.~\req{strongmaster} gives \BE
i\hbar\Delta R^{\a}+(-1)^{\e_{\b}}R^{\b}\d_{\b}R^{\a}-(X,R^{\a})=
(-1)^{\e_\a}(i\hbar\d_{\b}F^{\a\b} - \d_\b X F^{\a\b})\,,
\label{conditionfromstar}\EE which, in fact,  implies the consistency
condition \req{consistency}.  As is always the case with master equations
formulated with the appropriate boundary conditions at all antifields set to
zero, higher orders in the antifields generate higher compatibility
conditions.  However, the use of the $\l^*$-extended formalism is not limited
to putting together the compatibility conditions. It is very useful in
understanding gauge independence of the integral \req{pathintegral}.

\smallskip

In order to `legitimatize' the $\l^*$-extended formalism, consider the path
integral \req{pathintegral} rewritten as \BE\new\BA{rcl}
Z&=&\int\exp\biggl\{{i\over\hbar}\Bigl[W + \cX \Bigr]
\biggr\}\delta(\l^*)\rho(\Gamma)[d\Gamma][d\lambda][d\lambda^*]\\
&=&\int\exp\biggl\{{i\over\hbar}\Bigl[W + \cX + \l^*_\a\eta^\a \Bigr]
\biggr\}\rho(\Gamma)[d\Gamma][d\lambda][d\lambda^*][d\eta]
\EA\label{rewrite}\EE where we have used that $\cX\Bigm|_{\l^*=0}=X$. Now,
the point of rewriting \req{pathintegral} in this form was that $\cX$
satisfies a strong master equation. At the same time, we have represented the
partition function as an integral over the extended manifold $\tilde\cM$,
which bears the antibracket \req{Deltaextended}. However, \req{rewrite} is
obviously written in a very special hypergauge, namely the one that exlicitly
kills the $\l^*$-dependence. To make contact with the general gauge theory on
$\tilde\cM$ based on a solution to the strong master equation we have to show
that it is possible to choose arbitrary hypergauges in the integral
\req{rewrite}.  As before, we do that infinitesimally by performing a
non-constant shift of the integration variables and the accompanying
transformation of the form of \req{delta1Gamma}, {\it but now on the extended
manifold $\tilde\cM$\/}, using $(~,~)_{\rm ext}$ etc. The crucial observation
is that defining $\tilde W=W+\l^*_\a\eta^\a$ we would obtain a {\it
solution\/} of the master equation\label{bridge} \BE \half(\tilde W,\tilde
W)_{\rm ext}=i\hbar\Delta_{\rm ext}\tilde W\EE Hence we can repeat on
$\tilde\cM$ the steps \req{J1}--\req{delta1J} starting with
$\delta\tilde\Gamma=(\tilde\Gamma,\,-\tilde W + \cX)^{\phantom{Z}}_{\rm
ext}\mu$ (and $R^\a\to0$). We then arrive at a variation of $\cX$ which, as
will be seen shortly, describes the arbitrariness in $\cX$ corresponding to
the gauge freedom. We thus conclude that an arbitrary admissible gauge choice
is allowed in the integral \req{rewrite}, thereby providing the bridge
between the formalism of the previous subsection and the $\l^*$-extended
formalism.

\medskip

The advantage of introducing the $\l^*$-extended formalism is that, as
follows from \cite{BV-84,VT}, the automorphisms of the infinite algebra
generated by the `strong' master equation \req{strongmaster} can be described
in a closed form as
\BE\exp{\L(\frac{i}{\hbar}\cX'\R)}=\Bigl(\exp{\L[\Delta_{\rm ext},
\widetilde{\Psi}\R]}\Bigr)\exp{\L(\frac{i}{\hbar}\cX\R)}
\label{symplecticauto}\EE where $\widetilde{\Psi}$ is a fermionic function or
operator and $\Delta_{\rm ext}$ acts to the right (so that, e.$\;$g.,
$[\Delta_{\rm ext},\widetilde{\Psi}](F)= \Delta_{\rm
ext}(\widetilde{\Psi}\,F) +\widetilde{\Psi}\Delta_{\rm ext}F$).  Consider the
case when $\widetilde{\Psi}$ is an infinitesimal function $\widetilde{f}$.
Then, \BE \delta\cX\equiv\cX'-\cX=(\cX, \widetilde{f})_{\rm
ext}-i\hbar\Delta_{\rm ext}\widetilde{f}\label{delXf}\EE If we assume
$\widetilde{f}$ to be independent of $\l^*_{\a}$ we find ($\widetilde{f}=f$)
\BE \delta X=\L.\delta\cX\R|_{\l^*=0}=(X,f)-
(-1)^{\e_{\a}}R^{\a}\d_{\a}f-i\hbar\Delta f\label{adelX}\EE which is equal to
\req{deltaX}.  This shows that the variation $\delta X$ \req{deltaX} is
consistent since it agrees with the formula for infinitesimal automorphisms.
We also have \BE \delta R^\a = -\Bigl(\Dl{\lst_\a}\delta
\cX\Bigr)\biggm|_{\lst=0}= (R^\a, f)-(-1)^{\e_\b} F^{\b\a}\d_\b f
\label{deltaRfull}\EE which reproduces eq.~\req{deltaR} that has been
obtained as a part of the condition guaranteeing compatibility of the
`induced' variation \req{deltaX} with the weak master-equation \req{Xeq}.

The most general consistent variation is obtained from the automorphism
formula \req{symplecticauto} when the infinitesimal function $\widetilde{f}$
has an arbitrary dependence on $\l^*_\a$. Inserting \BE \widetilde f\equiv
f-\l^*_{\a}Q^{\a} +O((\l^*)^2)\label{dpsi}\EE into \req{delXf} one arrives at
\BE \delta X = \L.\delta\cX\R|_{\l^*=0} = (X, f)-i\hbar\Delta
 f-(-1)^{\e_{\a}}R^{\a}{\d}_{\a} f -X{\!\stackrel{\lea}{\d}}_{\a}Q^{\a}+
i\hbar(-1)^{\e_{\a}}{\d}_{\a}Q^{\a}\EE instead of \req{adelX}. This can be
obtained as an effective change of $X$ induced by the above manipulations in
the path integral \req{pathintegral} if one also adds the variation of
$\l^\a$ \BE \delta\l^{\a}=-Q^{\a}\,,\qquad Q^{\a}\equiv
Q^{\a}(\Gamma,\l;\hbar)\label{iltrans}\EE Notice that $\delta R^\a$ in this
case has a much more general form than \req{deltaRfull}.

We conclude that the partition function \req{pathintegral} is locally
independent of the gauge-fixing action $X$ provided equation \req{Xeq} for
the master-action $X$ is satisfied.  By a transformation of variables in the
integral we have induced a deformation of the gauge-fixing action $X$ that
preserves the equation imposed on $X$.

\medskip

\newcounter{remarks} \setcounter{remarks}{0} \noindent\underline{\sc
Remarks}\nopagebreak

\refstepcounter{remarks} \noindent{\bf \arabic{remarks}.}\label{canonical}
Note that the mapping \req{symplecticauto} amounts to an anti\/{\it
canonical\/} transformation. An important consequence is that it can
therefore be applied to {\it any\/} function, not necessarily the
master-action.

\smallskip

\refstepcounter{remarks} \noindent{\bf \arabic{remarks}.}\label{freedom} It
is possible to `abelianize' the hypergauge conditions $G_\a=0$, expressing
them in terms of a gauge fermion and a `rotation' matrix~\cite{BT94-2}. In
that case, one can trace how the independence of the solution of the master
equation from the corresponding automorphism transformations reformulates as
the arbitrariness in choosing the gauge fermion and the rotation matrix.
This gives the precise relation between the arbitrariness of solutions to the
master equation and the gauge freedom.

\subsection{On the structure of hypergauge-fixing actions\label{Planck}}\lvm
Let us stress that $X$ can in general be an arbitrary polynomial in $\l^\a$.
This corresponds to the possibility of introducing non-singular gauges such
as e.g.\ the $\a$-gauges in Yang--Mills. On the other hand, singular
hypergauges are those that explicitly involve the product of delta-functions
in the integrand.  Singular hypergauges result from taking $X$ to be linear
in $\l^\a$, $X=G_\a\l^\a$.  Hence $\l^{\a}$ are Lagrange multipliers for the
hypergauges $G_{\a}$, and integrating over $\l^\a$ in \req{pathintegral}
results in concentrating the integral on the locus $G_{\a}=0$.  For such
$X$'s, there is an important grading by the so called {\sl Planck number\/}
$\Pl$\@. This is defined as \cite{BT93-2}--\cite{BT94-2} \BE\new\BA{l}
\Pl(FG)=\Pl(F) + \Pl(G)\,,\qquad
\Pl(\Gamma^A)=0\,,\qquad\Pl(\hbar)=\Pl(\l^\a)=-\Pl(\lst_{\a})=1\,.\\
\EA\label{513}\EE The master-actions $X$ that lead to a singular hypergauge
are characterized by $\Pl(X)=1$. Then all the exponents in
\req{symplecticauto} have Planck number zero, while in \req{adelX}, e.g.\
$R^\a$ has Planck number 2.  A solution to  the set of equations derived by
expanding \req{Xeq} is given by \cite{BT94-1} (in the original notations,
which may now appear somewhat non-systematic, with
$G_\a\l^\a={\rm our\ }X_0$, $H={\rm our\ }X_1$) \begin{eqnarray}
X&=&G_{\a}\l^{\a}+i\hbar H\label{nsX}\\ R^{\a}&=&\half
U^{\a}_{\;\b\g}\l^{\g}\l^{\b}(-1)^{\e_{\b}}-i\hbar
V^{\a}_{\;\b}\l^{\b}-(i\hbar)^2\tilde{G}^{\a}\label{nsR} \end{eqnarray}
where the functions $G_{\a}, H,  U^{\a}_{\;\b\g}, V^{\a}_{\;\b}$ and
$\tilde{G}^{\a}$ satisfy\BE (G_{\a}, G_{\b})=G_{\g} U^{\g}_{\;\a\b}
\label{invG}\EE \BE (H, G_{\a})=\Delta G_{\a} -
U^{\b}_{\;\b\a}(-1)^{\e_{\b}}-G_{\b}V^{\b}_{\;\a}\EE \BE \Delta H-\half(H,
H) + V^{\a}_{\;\a}=G_{\a}\tilde{G}^{\a}\EE Next, $H$ is  what has been
considered in the previous papers on the subject as an additional measure
factor in the integral.  Eq.~\req{invG} is the classical starting point for
this solution and corresponds to \req{classXR}.  It says that the  gauge
functions $G_{\a}$ are in involution.

\section{Antitriplectic Lagrangian quantization\label{sec:4} }\lvm We now
return to the main subject of this paper, the triplectic quantization in the
geometrical setting developed in section~2.  As in the antisymplectic case
considered in the previous section, we first introduce an invariant action
$W(\Gamma;\hbar)$, which is now defined on a $6N$-dimensional antitriplectic
manifold $\cM$ and is required to satisfy the generalized master equations
with a `transport' term due to the $V^a$:
\BE\Bigl(\Delta^a+{i\over\hbar}V^a\Bigr)\exp\Bigl\{{i\over\hbar}W\Bigr\}=0
\label{3Wmaster}\EE or, equivalently, \BE\half(W,W)^a + V^aW=i\hbar\Delta^aW
\label{3Wmasterdown}\EE where $\Delta^a, V^a$ and the antibrackets are as
defined in section~\ref{sec:2} in arbitrary coordinates.
Eq.~\req{3Wmasterdown} is solved by an expansion in $\hbar$ with the
appropriate boundary conditions.

Recall that the triplectic quantization of a given theory requires
introducing not only antifields, but also a `second set of fields'.
Therefore, even with all the antifields set to zero, one has to single out a
submanifold $\cL_0$ of the original fields of the theory that is being
quantized. Most naively, one could think of the procedure of adding the new
fields as duplicating the original variables $\phi^\a\in\cL_0$,  that is
introducing, instead of the action $\cS(\phi)$, a function of two sets of
variables, $\cS(\phi^\a - \tilde\phi^\a)$, possessing the obvious `gauge'
symmetry.  However, the actual doubling of the degrees of freedom is not so
naive, and the procedure consists {\it not\/} in adding another copy of the
fields $\phi^\a$, but rather in going over to the cotangent bundle
$T^\ast\cL_0$ to the manifold $\cL_0$. The new `fields' are therefore certain
$\bar\phi_\a$, where the index position points to their `momentum' nature.

As to the boundary conditions on the master-action, one could therefore
declare that, after projecting out all the antifields, one should be left
with a cotangent bundle whose `coordinates' (as opposed to the `momenta') are
the original variables of the theory.  However, an obvious generalization,
which we are going to make, is achieved by replacing the cotangent bundle by
an arbitrary {\it symplectic\/} manifold and its Lagrangian
submanifold\footnote{Note that the Grassmann parity is {\it not\/} reversed
in this construction.  We will actually use only the Poisson structure, but
in order that the integrals be well-defined eventually, this must be
non-degenerate.}.  Then,  the boundary conditions are imposed on $W$ in two
steps.

To formulate the boundary conditions, one first restricts to a fixed
symplectic manifold $\cL_1\subset\cM$  with a chosen Lagrangian submanifold
$\cL_0\subset\cL_1$.  Then, one imposes \BE
W(\,\cdot\,;0)\Bigm|_{\cL_0}=\cS(\,\cdot\,)\EE Note that the dimensions are
given by the formulae \req{Ldimension} and \req{Mdimension} and \BE
\dim\cL_0=(n|N-n)\,. \EE The submanifolds $\cL_0\subset\cL_1\subset\cM$ are
fixed and make up a part of the definition of the theory.  The `classical'
gauge generators are encoded in the boundary conditions involving $W^{(1)}$,
which is formulated as the appropriate symplectic analogue of \req{BC1}.

\medskip

Now we propose the following ansatz for the partition function path integral
in the triplectic quantization:  \BE Z=\int\exp\biggl\{{i\over\hbar}\Bigl[W +
X\Bigr] \biggr\} \rho(\Gamma)[d\Gamma][d\lambda]\label{3pathintegral}\EE
where $X(\Gamma,\l;\hbar)$ is a  gauge-fixing action. As before, $\l^{\a}$,
$\a=1,\ldots,N$, are parametric variables that generalize Lagrange
multipliers for the hypergauge functions.  We require that
\req{3pathintegral} be invariant under a generalized canonical transformation
accompanied by a shift of the $\l^\a$:  \BE\new\BA{rcl}
\delta\Gamma^A&=&{}(\Gamma^A, -W+X)^a\mu_a - 2V^{A a}\mu_a\,,\\
\delta\l^{\a}&=&{}-2R^{\a a}\mu_a \EA\label{Bltransform}\EE where $\mu_a$
are two fermionic constants.  This allows us to arrive at an equation we will
then postulate for $X$.  The form of \req{Bltransform} is such that the
resulting conditions on $X$ will not involve $W$ when one makes use of the
fact that $W$ satisfies the generalized quantum master equation
\req{3Wmasterdown}.  We claim that the integral \req{3pathintegral} will be
invariant under the transformations \req{Bltransform} provided \BE \half(X,
X)^a-i\hbar\Delta^a X-V^aX-X{\!\stackrel{\lea}{\d}}_{\a}R^{\a a}+ i\hbar
R^{\a a}{\!\stackrel{\lea}{\d}}_{\a}= 0\label{3Xeq}\EE or equivalently \BE
\Bigl(\Delta^a-\frac{i}{\hbar}V^a- \frac{i}{\hbar}(-1)^{\e_{\a}}R^{\a
a}\d_{\a}+ (-1)^{\e_{\a}}\d_{\a}R^{\a a}\Bigr)\exp{\L(\frac{i}{\hbar}X\R)}=0
\label{3Xeqexp}\EE where as before $\d_{\a}\equiv\d/\d\l^{\a}$ and
$\e_{\a}\equiv\e(\l^{\a})$.  This is a `weak' analogue of the generalized
quantum master equation \req{3Wmasterdown}, with an opposite sign for the
$V^a$-term.  As in the \asy\ case, solutions to \req{3Xeq} should be looked
for in the form of (formal) power series $X(\Gamma,\l;\hbar)$ and $R^{\a
a}(\Gamma,\l;\hbar)$ in $\hbar$:  \BE\new\BA{rcl} X&=&X_0+i\hbar
X_1+(i\hbar)^2X_2+\ldots\,,\\ R^{\a a}&=&R^{\a a}_0+i\hbar R^{\a
a}_1+(i\hbar)^2R^{\a a}_2+\ldots\label{3expansion} \EA\EE In the lowest
order, in particular, we find \BE \half(X_0,
X_0)^a-V^aX_0-X_0{\!\stackrel{\lea}{\d}}_{\a}R_0^{\a a}=0\label{classsXeq}
\EE As before, coefficients for all powers of $\l^{\a}$  must agree.

\smallskip

By applying to the LHS of eq.~\req{3Xeq} the operator
\BE\Delta^b-\frac{i}{\hbar}V^b+\frac{i}{\hbar}(X,\,\cdot\,)^b-
\frac{i}{\hbar}(-1)^{\e_\a}R^{\a b}\d_\a\EE and symmetrizing in $a$ and $b$
we arrive at the consistency condition
\BE\Dl{\l^\a}\L\{\L(i\hbar(\Delta^{\{a}R^{\a b\}})+V^{\{a}R^{\a b\}}-(X,
R^{\a\{ a})^{b\}}+(-1)^{\e_\b}R^{\b \{a}\d_\b R^{\a b\}}\R)
e^{\frac{i}{\hbar}X}\R\}=0\label{abconsistency}\EE which is solved in the
extended formalism introduced below, see \req{Rsolution}.

\smallskip

\section{Independence from the choice of hypergauge fixing and \hfill\break
gauge invariance in the antitriplectic case\label{sec:5}}\lvm In this section
we shall show that the path integral \req{3pathintegral} is independent of
local variations of hypergauge-fixing action $X(\Gamma,\l;\hbar)$ for a
certain class of variations.  This actually amounts to the statement of gauge
invariance in the sense of independence from the {\it ordinary\/}
gauge-fixing function.  In the geometrically covariant version, this function
is encoded into a set of hypergauge-fixing functions which satisfy master
equations; thus gauge invariance takes the form of an {\it adequate\/}
freedom in choosing hypergauge-fixing functions.  The fundamental fact is
that this freedom is determined just by the (master) equation satisfied by
the gauge-fixing master-action $X$.  Therefore gauge invariance is
reformulated in terms of automorphisms of solutions to the master equation.

We start with a procedure that is a generalization of the previous approach
in the antisymplectic case given in section~\ref{sec:3}, \ie we shall look
for a transformation of integration variables that can induce a variation of
$X$.  Let us replace the fermionic constants $\mu_a$ in \req{Bltransform} by
arbitrary functions of $\Gamma^A, \l^{\a}$ and $\hbar$ subjected to the
condition that they are  expandable in powers of both $\l^{\a}$ and $\hbar$.
The integrand \req{3pathintegral} is then no longer invariant. In fact, the
transformations \req{Bltransform} give now rise to the following extra
Jacobian $1+\delta J$ in \req{3pathintegral}:  \BE \delta J=(-W+X,
\mu_a)^a-2V^a\mu_a-2R^{\a a}{\d}_{\a}\mu_a(-1)^{\e_{\a}}\EE (The second term
on the RHS is the action of vector fields on the respective functions; we
also remind the reader that $\d_\a=\d/\d\l^\a$).  The contribution $(-W,
\mu_a)^a$ may  be compensated for by the additional transformation \BE
\delta_1\Gamma^A=\frac{\hbar}{i}(\Gamma^A, \mu_a)^a\EE which in turn yields
a new Jacobian \BE \delta_1J=2\Delta^a\mu_a\EE and at the same time doubles
the term $(X,\mu_a)^a$. In this way we finally obtain the following effective
change of the integrand of \req{3pathintegral}:  \BE \delta\L[\rm exponent\
of\ \req{3pathintegral}\R] =\frac{i}{\hbar}\delta X\EE where \BE \delta
X=(X, f_a)^a-V^a f_a-i\hbar\Delta^a f_a-(-1)^{\e_{\a}}R^{\a a}\d_{\a} f_a
\label{3deltaX}\EE with \BE f_a\equiv\frac{2\hbar}{i}\mu_a\,.\EE We may also
add a further variation of $\l^\a$ of the form \req{iltrans} which results in
the following effective change of $X$:  \BE \delta X = (X,  f_a)^a-V^a f_a-
(-1)^{\e_{\a}}R^{\a a}\d_{\a} f_a - X{\!\stackrel{\lea}{\d}}_\a Q^{\a}
-i\hbar\Delta^a f_a + i\hbar(-1)^{\e_\a} {\d}_{\a}
Q^{\a}\,.\label{3deltaXgeneral}\EE

Now, as in the \asy\ case, we have to ensure that the master equation imposed
on $X$ will be preserved after the deformation.  However, not every $\delta
X$ given by \req{3deltaXgeneral} would preserve the triplectic master
equations (recall that these include the algebra of gauge conditions as well
as other relations). The point is that the RHS of \req{3deltaXgeneral}
involves {\it two\/} fermionic functions $f_a$, $a=1,2$ whereas, as we will
see shortly, a natural arbitrariness in solutions to the master equation, as
well as the gauge freedom, is described by {\it one\/} bosonic function.  We
can nevertheless show that there do exist certain variations of the form
\req{3deltaXgeneral} that preserve the master equation.  This will follow
from the extended formalism, which is constructed as follows.

We introduce a linear space $\Lambda$ spanned by $(\l^\a, \l^*_{\a a},
\bar\l_\a, \eta^{\a a})$, with
$\epsilon(\l^\a)=\epsilon(\bar\l_\a)=\epsilon_\a$, $\epsilon(\l^*_{\a
a})=\epsilon(\eta^{\a a})=\epsilon_\a+1$, and define an extended triplectic
manifold $\tilde\cM=\cM\times\Lambda$\@. On $\tilde\cM$, we introduce the
operators \BE \Delta^a_{\rm ext}=
\Delta^a+(-1)^{\e_\a}\Dl{\l^\a}\Dl{\l^\ast_{\a a}} +
(-1)^{\e_\a+1}\epsilon^{ab}\Dl{\bar\l_\a}\Dl{\eta^{\a b}}\label{Deltaext}
\EE and the corresponding antibrackets \BE (F,G)^a_{\rm ext}= (F,G)^a +
\Bigl(\Dr{\l^\a}{F}\Dl{\lst_{\a a}}{G} + \epsilon^{ab}\Dr{\eta^{\a
b}}{F}\Dl{\bar\l_\a}{G} -(-1)^{(\e(F)+1)(\e(G)+1)}(F\leftrightarrow
G)\Bigr)\,. \EE Another triplectic quantity are the vector fields $V^a$,
which extend to $\tilde\cM$ as \BE \cV^a = V^a - \half\Bigl(
\epsilon^{ab}\l^\ast_{\a b}\Dl{\bar\l_\a} - (-1)^{\e_\a}\eta^{\a
a}\Dl{\l^\a}\Bigr)\label{Vext}\EE and satisfy the necessary conditions
$[\cV^a,\,\Delta^b_{\rm ext}]=0$ etc.\ in accordance with the conditions
given in section~\ref{sec:2}. (Notice that equations
\req{Deltaext}--\req{Vext} follow from the Darboux expressions
\req{DeltaDarboux}--\req{Vflat}.)

Now we can introduce an extended quantum master equation \BE \L(\Delta^a_{\rm
ext}-\frac{i}{\hbar}\cV^a\R)\exp{\L(\frac{i}{\hbar}\cX\R)}=0\label{semaster}
\EE or equivalently, \BE\half(\cX,\,\cX)^a_{\rm
ext}-\cV^a\cX-i\hbar\Delta^a_{\rm ext}\cX=0\,.\label{strongextended}\EE
The `weak' master equation \req{3Xeqexp} follows from \req{strongextended}
once we take $\cX$ to be \BE \cX(\Gamma, \l, \l^*, \bar\l, \eta) =
\tX(\Gamma, \l, \l^*, \bar\l) + \half\l^*_{\a a}\eta^{\a a}\label{lineareta}
\EE and further expand $\tX$ as \BE \tX(\Gamma, \l, \l^*, \bar\l) = X(\Gamma,
\l) - \l^*_{\a a}R^{\a a}(\Gamma, \l) - \bar\l_\a\bar R^\a(\Gamma, \l) + {\rm
higher\ orders\ in\ }\l^*,\bar\l\label{3starexpansion}\EE (That the
dependence of $\cX$ on $\eta$ is linear can also be understood by reducing
from the level-1 theory).  Further, specifying the second-order terms in
\req{3starexpansion} as \BE \half\lst_{\a a}\lst_{\b b} F^{\a a;\b b} +
\half\bar\l_\a\bar\l_\b \bar F^{\a\b} + \bar\l_\b\lst_{\a a}E^{\a a;\b}\EE
(with $F^{\b b;\a a}=(-1)^{(\e_{\a}+1)(\e_{\b}+1)}F^{\a a;\b b}$), we find
from the first order in $\lst$ in eq.~\req{strongextended}:  \BE
i\hbar\Delta^a R^{\a b} + V^aR^{\a b} + (-1)^{\e_\b}R^{\b a}\d_\b R^{\a b} -
(X,\,R^{\a b})^a + (-1)^{\e_\a}\epsilon^{ab}\bar R^\a =
(-1)^{\e_\b}(i\hbar\d_\b F^{\a b;\b a} - \d_\b X F^{\a b;\b a})
\label{Rsolution}\EE This solves consistency condition \req{abconsistency}
upon symmetrization, while $\bar R^\a$ represents the antisymmetrized part.
{}From the first order in $\bar\l$  in eq.~\req{strongextended} we get further
relations constraining $\bar R^\a$:  \BE i\hbar\Delta^a\bar R^\a + V^a\bar
R^\a + (-1)^{\e_\b}R^{\b a}\d_\b\bar R^\a - (X,\,\bar R^\a)^a =
(-1)^{\e_\b}(i\hbar\d_\b E^{\b a;\a} - \d_\b X E^{\b a;\a})\,. \EE

\smallskip

Let us show now how the integral \req{3pathintegral} can be reformulated on
the extended triplectic manifold using the extended master action $\cX$. We
have \BE\new\BA{rcl} Z&=&\int\exp\biggl\{{i\over\hbar}\Bigl[W + \tX
\Bigr]\biggr\} \delta(\l^*)\delta(\bar\l)
\rho(\Gamma)[d\Gamma][d\lambda][d\lambda^*][d\bar\lambda]\\
&=&\int\exp\biggl\{{i\over\hbar} \Bigl[W + \tX + \l^*_{\a a}\eta^{\a a} +
\bar\l_\a\xi^\a \Bigr] \biggr\}\rho(\Gamma)
\underbrace{[d\Gamma][d\lambda][d\lambda^*]
[d\bar\lambda][d\eta]}_{[d\tilde\Gamma]}[d\xi]\\
&=&\int\exp\biggl\{{i\over\hbar} \Bigl[\cW + \cX \Bigr]
\biggr\}\rho(\tilde\Gamma)[d\tilde\Gamma][d\xi] \EA\label{3rewrite}\EE
($\xi\equiv\l^{(1)}$ will become Lagrange multipliers of the next-{\it
level\/} theory), where $\cW=W+\half\l^*_{\a a}\eta^{\a a} + \bar\l_\a\xi^\a$
and $\cX=\tX+\half\l^*_{\a a}\eta^{\a a}$ are {\it solutions\/} on
$\tilde\cM$ to the strong master equations:  \BE\new\BA{rcccl}
\half(\cX,\,\cX)^a_{\rm ext} &-& \cV^a\cX &=&i\hbar\Delta^a_{\rm ext}\cX\\
\half(\cW,\,\cW)^a_{\rm ext} &+& \cV^a\cW &=&i\hbar\Delta^a_{\rm ext}\cW
\EA\EE Therefore, as in section~\ref{sec:3}, we can perform an arbitrary
gauge variation in the integral \req{3rewrite}, thus establishing the
validity of the path integral on $\tilde\cM$ in a general hypergauge.  We
also have at our disposal now all the power of the automorphism formula for
solutions of the strong master equations. This is described as follows.

Consider eq.~\req{semaster}.  We have a mapping $\cX\mapsto\cX'$ on the
solutions given by \cite{BLT-4} \BE \exp{\L(\frac{i}{\hbar}\cX'\R)}=\L(\exp
\epsilon_{ab}\frac{\hbar}{i}[\Delta^a_{\rm ext}-\frac{i}{\hbar}\cV^a,\,
[\Delta^b_{\rm ext}-\frac{i}{\hbar}\cV^b,\tPhi]\,]\R)
\exp{\L(\frac{i}{\hbar}\cX\R)}\label{aut}\EE where $\tPhi$ is an arbitrary
{\em bosonic} function or operator.  When $\tPhi$ is an infinitesimal
function $\tilde\varphi$, we have \BE\new\BA{rcl} \delta\cX\equiv\cX'-\cX
&=&\cK\tilde\varphi+ \epsilon_{ab}\Bigl\{(\cX,\,(\cX,\,\tilde\varphi)^b_{\rm
ext})^a_{\rm ext} -\cV^a(\cX,\,\tilde\varphi)^b_{\rm ext}-
(\cX,\,\cV^b\tilde\varphi)^a_{\rm ext} \\
{}&{}&{}+i\hbar\Bigl[(\cV^a\Delta^b_{\rm ext}+\Delta^a_{\rm
ext}\cV^b)\tilde\varphi-(\cX, \Delta^b_{\rm ext}\tilde\varphi)^a_{\rm ext}-
\Delta^a_{\rm ext}(\cX, \tilde\varphi)^b_{\rm ext}\Bigr] \\{}&{}&{}-
\hbar^2\Delta^a_{\rm ext}\Delta^b_{\rm ext}\tilde\varphi\Bigr\}
\EA\label{deltatX}\EE (where $\cK=\epsilon_{ab}\cV^a\cV^b$).  This expression
rewrites as an $\cX$-dependent transformation \BE \delta\cX= (\cX,
\widetilde{f}_a)^a_{\rm ext}-\cV^a\widetilde{f}_a -i\hbar\Delta^a_{\rm
ext}\widetilde{f}_a\label{first}\EE where \BE
\widetilde{f}_a=\epsilon_{ab}\L\{(\cX, \tilde\varphi)^b_{\rm
ext}-\cV^b\tilde\varphi- i\hbar\Delta^b_{\rm ext} \tilde\varphi\R\}\,.
\label{second}\EE The form \req{lineareta} of $\cX$ depending on $\eta$ only
via $\half\l^*_{\a a}\eta^{\a a}$ will be preserved once $\tilde\varphi$ is
taken to be a function of $(\Gamma,\l,\l^*,\bar\l)$. Then, the variation of
the master action $X$ follows as $\delta X=\delta\cX\Bigm|_{\l^*=0,\bar\l=0}$
and turns out to have the form of a variation \req{3deltaXgeneral} obtained
above.  {}From \req{3starexpansion} and \req{deltatX}, we also get the
variation of $R^{\a a}$ as \BE\delta R^{\a a}= -\L(\frac{\d}{\d\l^*_{\a
a}}\delta\cX\R)\biggm|_{\l^*=0,\,\bar\l=0}\label{3deltaR}\EE

Thus the automorphisms of the strong master-equation in the extended sector
reduce to the original phase space as a particular realization (certain
$X$-dependent transformations) of the variations \req{3deltaXgeneral} that
were obtained by changing variables in the path integral.  The partition
function \req{3pathintegral} is therefore independent of local changes of the
gauge-fixing master action $X$ within the class of those $X$'s which satisfy
the master equation; in other words, the subclass of the variations
\req{3deltaXgeneral} consisting of the above $X$-dependent transformations do
preserve the algebra of the constraints imposed by the master-action $X$.

To assert that the above transformations of $X$ describe gauge invariance of
our formalism, one needs the statement that the automorphism formula
\req{aut} describes the {\it maximum\/} freedom allowed by (generic)
solutions to the master equation. As we have noted, in the \asy\ case a
similar statement can be proven using `abelianization', whose triplectic
analogue has yet to be elaborated. It seems very probable, however, that the
above automorphism formula describes correctly the maximal arbitrariness of a
solution to the master equation, thereby allowing one to arrive at an {\it
arbitrary\/} gauge (in the ordinary  sense) by choosing different solutions
to the master equation.

\medskip

\newcounter{remarks5} \setcounter{remarks5}{0} \noindent\underline{\sc
Remarks.}\nopagebreak

\refstepcounter{remarks5} \noindent{\bf \arabic{remarks5}.}\label{coordinate}
In similarity with the \asy\ case, we can find a {\it coordinate
transformation\/} that induces the above variation when applied to $X$; the
power of the `coordinate representation' of this variation is that it can
then be applied to {\it any\/} function on $\cM$.  It turns out that the
desired infinitesimal diffeomorphism on $\cM$ consists of a `hamiltonian'
piece (in the triplectic sense, i.e.\ generated by the antibrackets) and a
term involving $V^a$.  The `hamiltonian', or canonical, transformations are
called so by analogy with the (anti)symplectic formalism; they are considered
systematically in Appendix~B\@. While a naive generalization of the canonical
transformation to the triplectic case would be given by $F\mapsto(h_1, F)^1 +
(h_2, F)^2$ which involves {\it two\/} `hamiltonians' $h_a$, the intrinsic
triplectic transformations are those for which the (fermionic) functions
$h_a$ are expressed through one arbitrary bosonic function $\varphi$ and a
{\it solution\/} $X$ of the master equation. One thus gets $X$-dependent
transformations, which are a characteristic feature of the triplectic
geometry.

To describe how the transformation generated by $\varphi$ acts on an
arbitrary function $F$, one should start with the extended formalism.
Consider the following infinitesimal mapping on functions on the extended
manifold $\tilde\cM$:  \BE\widetilde F\mapsto\widetilde F
+\epsilon_{ab}(\widetilde F,\,(\cX,\, \widetilde \varphi)_{\rm ext}^b)_{\rm
ext}^a + 2\epsilon_{ab}(\widetilde F,\,\cV^a\widetilde \varphi)_{\rm ext}^b
-\epsilon_{ab}(\cV^a\widetilde F,\,\widetilde \varphi)_{\rm ext}^b
\label{coordinatemapext}\EE where different $\widetilde\varphi$'s label
infinitesimal automorphisms and $\cX$ is a fixed solution of the master
equation.  When applied to $\widetilde F=\cX$, this transformations
reproduces the classical ($\hbar$-independent) part of eq.~\req{deltatX}
apart from the $\cK\tilde\varphi$ term.  Now, by expansion in $\l^*,\bar\l$,
one arrives at a `weak' form of the corresponding transformation.  Its `main'
part will still be given by antibrackets and $V^a$ fields. To avoid
overloading our formulae with the details that bear no principal importance,
we will write out only this `main' part of the coordinate transformation. It
is obtained by taking the fields in \req{coordinatemapext} to depend on
$(\Gamma, \l)$ only.  Thus, consider an infinitesimal variation of functions
$F$ on $\cM$:  \BE F\mapsto F+\epsilon_{ab}(F,\,(X,\,\varphi)^b)^a +
2\epsilon_{ab}(F,\,V^a\varphi)^b -\epsilon_{ab}(V^aF,\,\varphi)^b
\label{coordinatemap}\EE When applied to $X$, this results in the following
transformation \BE\delta X=(X,  f_a)^a-V^a f_a-i\hbar\Delta^a f_a\,,
\label{241}\EE with $ f_a$ in turn depending on $X$ via \BE
f_a=\epsilon_{ab}\left\{(X, \varphi)^b-
V^b\varphi-i\hbar\Delta^b\varphi\right\}\,.\label{242}\EE

The first two terms in the transformation \req{coordinatemap} are given by a
hamiltonian vector field $\oV^a_{h_a(\varphi)}$ (see \req{oVa}) with the
hamiltonians \BE h_a(\varphi) = -\epsilon_{ab}(X,\,\varphi)^b +
2\epsilon_{ab}V^b\varphi\,,\label{hamiltonian}\EE while the third term in
\req{coordinatemap} is a `transport' term.  Thus \req{coordinatemap} rewrites
as \BE \delta F=\oV^a_{h_a(\varphi)}F - \epsilon_{ab}(V^aF,\,\varphi)^b\,.
\label{coordinatemapH}\EE This becomes purely `hamiltonian' if $V^aF=0$, an
important example of which we will meet in section~\ref{sec:6}.

\smallskip

\refstepcounter{remarks5} \noindent{\bf \arabic{remarks5}.}\label{direct}
Some consequences of the automorphism formula, such as the invariance of the
strong master equation under variations given by \req{first}, \req{second},
can also be verified independently \cite{BM}:  Consider varying the extended
`strong' master-equation \BE(\cX,\delta \cX)_{\rm ext}^a-\cV^a\delta\cX=
i\hbar\Delta_{\rm ext}^a\delta\cX\label{23}\EE where $\delta \cX$ is given by
\req{first}, \req{second}.  In order to demonstrate that  this variation
indeed satisfies \req{23} one needs the following identities (we suppress the
index `ext') \cite{BM}:  \BE \epsilon_{bc}(\cX, (\cX, (\cX,
\tilde\varphi)^c)^b)^a= \epsilon_{bc}\left\{(\cX, (\tilde\varphi, (\cX,
\cX)^b)^c)^a +(\tilde\varphi, (\cX, (\cX, \cX)^b)^c)^a+2((\cX, \cX)^b, (\cX,
\tilde\varphi)^c)^a\right\}\label{34}\EE \BE\new\BA{rcl}
\epsilon_{bc}\left\{\Delta^a(\cX, (\cX, \tilde\varphi)^c)^b+ (\tilde\varphi,
(\cX, \Delta^c \cX)^b)^a+(\Delta^c\cX,(\cX,\tilde\varphi)^b)^a-(\cX,
(\Delta^c\cX,\tilde\varphi)^b)^a+\right.\\
\left.{}+(\cX,(\cX,\Delta^c\tilde\varphi)^b)^a+(\cX,
\Delta^b(\cX,\tilde\varphi)^c)^a-(\Delta^b \cX, (\cX, \tilde\varphi)^c)^a
\right\} =\\ =\epsilon_{bc}\half\left\{\Delta^a(\tilde\varphi,
(\cX,\cX)^b)^c+(\tilde\varphi,\Delta^c(\cX,
\cX)^b)^a+2(\Delta^b\tilde\varphi, (\cX, \cX)^c)^a \right\}\EA\label{35}\EE
and \BE\new\BA{l} \epsilon_{bc}\Bigl\{\Delta^a\Delta^b(\cX, \tilde\varphi)^c-
\Delta^a(\Delta^c\cX, \tilde\varphi)^b\\ {}+ 2(\Delta^c\cX,
\Delta^b\tilde\varphi)^a +(\tilde\varphi, \Delta^b\Delta^c\cX)^a+
\Delta^a(\cX, \Delta^c\tilde\varphi)^b+(\cX, \Delta^b\Delta^c\tilde\varphi)^a
\Bigr\} = 0\EA\label{36}\EE They can be derived using \req{DeltaFG} and the
identities $\Delta^a\Delta^b\Delta^c(\cX^3\tilde\varphi)=0$,
$\Delta^a\Delta^b \Delta^c(\cX^2\tilde\varphi)= 0$ and
$\Delta^a\Delta^b\Delta^c(\cX\tilde\varphi)= 0$ respectively. One may notice
that these identities are even valid when $\Delta^a_{\rm ext}$ is replaced by
$\Delta_{\rm ext}^a+\a\cV^a$ for any constant $\a$ (cf.~\req{Deltapm}!).

\smallskip

\refstepcounter{remarks5} \noindent{\bf
\arabic{remarks5}.}\label{divergence5} We have remarked in
section~\ref{sec:2} that the condition $\div V^a=0$ can be removed at the
expense of several new terms appearing in some of our formulae.  Namely, the
master equation \req{strongextended} will change to \BE
\half(\cX,\,\cX)^a_{\rm ext}-\cV^a\cX-i\hbar\Delta^a_{\rm ext}\cX +\half
i\hbar\div V^a=0\EE (recall that $\div V^a=\rho^{-1}\d_A(\rho V^{A
a})(-1)^{\e_A}$), which suggests introducing a differential operator
\BE\hV^a=\cV^a+\half\div V^a\EE (as $\Lambda$ is a linear space, it does not
contribute to the density $\rho$; divergence of the components of $\cV$ along
$\Lambda$ is zero).  We still have to ensure
$[\hV^a,\,\hV^b]=0\Leftrightarrow\hV^{\{a}\hV^{b\}} =0$ and $[\Delta_{\rm
ext}^a,\,\hV^b]=0$. This reduces to requirements on the $V^A$ components, and
results in the following modifications in the formulae in
section~\ref{sec:2}. First, in addition to \req{VVcoord} we are going to have
\BE (-1)^{\e_A}V^{A a}\d_A\Bigl(\rho^{-1}\d_B\bigl(\rho V^{Bb}\bigr)
(-1)^{\e_B}\Bigr)=0\,. \EE Eq.~\req{VEformula} does not change, while
\req{VEformula2} acquires on the LHS an additional term \BE
E^{ABa}\d_A\Bigl(\rho^{-1}\d_B\bigl(\rho V^{Bb}\bigr) (-1)^{\e_B}\Bigr)\,.
\EE Finally, new equations are \BE \Delta^a\Bigl(\rho^{-1}\d_B\bigl(\rho
V^{Bb}\bigr) (-1)^{\e_B}\Bigr)=0\,. \EE Note also that the master equation
\req{3Wmasterdown} changes accordingly, into \BE
\Bigl(\Delta^a+{i\over\hbar}\hV^a\Bigr)\exp\Bigl\{{i\over\hbar}W\Bigr\}=0\,.
\EE

\smallskip

\refstepcounter{remarks5} \noindent{\bf \arabic{remarks5}.}\label{Poisson5}
Note finally that the Poisson bracket extends to $\cL_1\times\Lambda_1$,
where $\Lambda_1$ is spanned by $(\l^\a, \bar\l_\a)$.

\section{On the structure of hypergauge-fixing actions\label{sec:6}}\lvm A
useful illustration of the general scheme is provided by taking $X_0$, the
classical part of the gauge-fixing master-action $X$, to depend on $\l^\a$ at
most linearly, as \BE X_0=G_\a\l^\a +Z\label{X0}\EE where $Z$ is independent
of $\lambda^\a$.  Then the functions $G_\a$ should eventually specify a
Lagrangian submanifold and thereby fix a gauge, since integrating over
$\l^\a$ introduces delta functions $\prod_\a\delta(G_\a)$ into the integral
\req{3pathintegral}. This makes the form \req{X0} important in applications.

It follows that for master actions of this class one can define a `weakened'
analogue of the Planck number introduced in the \asy\ case in
section~\ref{Planck}. In the triplectic case, however, we will have a {\it
filtration\/} instead of a grading.  First, we mimic \req{513}:
\BE\new\BA{l} \Pl(FG)=\Pl(F) + \Pl(G)\,,\\ \Pl(\Gamma^A)=0\,, \\
\Pl(\hbar)=\Pl(\l^\a)=-\Pl(\lst_{\a a})=1\,,\\
\Pl(\bar\lambda_\a)=-\Pl(\eta^{\a a})=-2\,.  \EA\EE However, by assigning a
Planck number $p$ to a function we will now mean that the function can
contain all terms whose individual Planck numbers are not bigger than $p$.
Then, the ansatz \req{X0} is characterized by Planck number 1. Such $X$'s
provide a natural triplectic analogue of singular gauges (i.e.\ those
resulting in an explicit insertion of delta-functions into the integrand in
\req{3pathintegral}).

For the classical part of the master equation, we find from
eq.~\req{classsXeq} \BE \half(X_0,X_0)^a - V^aX_0 = G_\a R^{\a a}_0
\label{XmasterM}\EE with structure functions $R^{\a a}_0$ (see
\req{3expansion}) that must now have the form \BE R^{\a a}_0=\half U^{\a
a}_{\b\g}\l^\g\l^\b(-1)^{\e_\b} + U^{\a a}_\b\l^\b + \half U^{\a a}
\label{M}\EE for some $\l^\a$-independent $U^{\a a}_{\b\g}$, $U^{\a a}_\b$,
and $U^{\a a}$.  Expanding in $\l^\a$, one readily extracts the involution
relations of the hypergauge functions $G_\a$ \BE (G_\a,G_\b)^a=G_\g U^{\g
a}_{\a\b}\label{GGinvolution}\label{37}\EE with structure functions $U^{\g
a}_{\a\b}$, together with \BE (Z, G_\a)^a - V^aG_\a =G_\b U^{\b a}_\a\,,
\label{38}\EE and \BE (Z, Z)^a=G_\g U^{\g a}\,.\label{39}\EE The involution
relation \req{GGinvolution} is a straightforward generalization of what we
had in the \asy\ case. However, there are no counterparts to
eqs.~\req{38}--\req{39}. This is not yet the final form of the equation:
below, we will propose a further specialization of the form of $Z$.

Going over to higher-order terms in the expansion of $X$ in powers of $\hbar$
beyond the zeroth order given by \req{X0}, we extend the ansatz~\req{X0},
which is a first-order polynomial in $\l^\a$, to an $X$ of {\it Planck number
1\/}. This means in particular that $X_2$ etc.\ ${}=0$, while $X_1$ is
$\l$-independent.  We then arrive at  the following equations for the
`measure' $H\equiv X_1$:  \BE\new\BA{l} (H,G_\a)^a = \Delta^aG_\a  - U^{\b
a}_{\b\a}(-1)^{\e_\b} + G_\b P^{\b a}_\a\,,\\ (H,Z)^a - V^a H = \Delta^a Z  -
U^{\a a}_{\a} + G_\a P^{\a a}\,,\\ \Delta^a H-\half(H,H)^a =  P^{\a a}_{\a} -
G_\a R_2^{\a a} \EA\label{hbarorder}\EE which corresponds to $R_1^{\a a}$
(see \req{3expansion} and \req{M}) having the structure \BE R_1^{\a a}=P^{\a
a}_\b\l^\b+P^{\a a}\,,\EE with $R_2^{\a a}$ being $\l$-independent. The two
terms in the $\hbar$-expansion can be called the `classical' master-action
and the triplectic measure\footnote{It appears rather non-trivial that a
solution exists for the triplectic measure that has no $\hbar^2$-terms and no
$\l$ in the $\hbar$ order. In the \asy\ case, a similar solution is given by
a subtle differential-geometric construction \cite{BT94-2,S,Kh}, which would
be interesting to extend to the triplectic case.}.

\smallskip

Once we have chosen an $X_0$ having Planck number 1, we must ensure that the
variation of $X$ does not increase the Planck number. To this end, we should
extract from \req{deltatX} terms of (individual) Planck number 2 (denoted
below as ${}|^{(2)}$), and require their vanishing.  We replace $\cV^a\cX$ in
\req{deltatX} using \req{strongextended}; then we observe that
$\widetilde\varphi$ must have Planck number zero in order to avoid terms with
yet higher Planck numbers.  Then the condition for the vanishing of (strict-)
Planck-number-2 terms becomes \BE
\epsilon_{ab}\L\{(\cX,(\cX,\tilde\varphi)^b_{\rm ext})^a_{\rm
ext}\Bigm|^{(2)} -\half((\cX,\,\cX)^a,\,\tilde\varphi)^b_{\rm
ext}\Bigm|^{(2)} +i\hbar(\Delta^a_{\rm ext}\cX,\,\tilde\varphi)^b_{\rm
ext}\Bigm|^{(2)} -\hbar^2\Delta^a_{\rm ext}\Delta^b_{\rm
ext}\tilde\varphi\Bigm|^{(2)}\R\}=0\label{Planck1general}\EE Expanding this
in $\l^*\bar,\l$ and $\hbar$ gives a set of relations guaranteeing
preservation of the ansatz \req{X0} under variations.  In the classical part,
the corresponding condition on $\varphi$ amounts to demanding the
cancellation of $\l$-squared terms.  This leaves us with the following
constraint on (the classical part of)
$\varphi=\widetilde\varphi|_{\l^*=0,\bar\l=0}$ and $G_\a$, $G_\b$ (which all
are functions on $\cM$):  \BE \epsilon_{ab}(G_\a,(G_\b,\varphi)^b)^a +
\epsilon_{ab}((G_\a,\varphi)^b,G_\b)^a(-1)^{\e_\b+1}=
\epsilon_{ab}((G_\a,G_\b)^a,\varphi)^b\label{skewJacobi}\EE which happens to
coincide with the Jacobi identities -- for the particular entries involved --
in the antisymmetric sector (in $ab$).  This is rather natural, since the
fact that the $\l$'s retain their r\^ole of Lagrange multipliers (setting the
integral onto the locus $G_\a\!=\!0$) means, by consistency, that the
antibracket algebra of the gauge functions $G_\a$ must be preserved.  For
general transformations this is impossible just because of the lack of the
Jacobi identity for two non-coincident antibrackets; only the compatibility
condition \req{newJacobi} symmetrized in $ab$ exists in general.

\smallskip

We will now further specialize the ansatz for $X_0$ to \BE X_0=G_\a\l^\a +\sK
Y\label{X00}\EE where $Y$ is a function on $\cM$ and $\sK$ was introduced in
\req{cK}.  The $\sK Y$ form of the $\l$-independent term does give a solution
of the master equation in the Darboux coordinates \cite{BM} and, due to its
invariant form, it can be taken over to the general case. Then, however, we
have to make sure that this ansatz is stable under the `gauge' variations
that follow from \req{first}, \req{second}.  We discuss this issue in
Appendix~B, where we also consider, more generally, {\it form\/} variations
of the triplectic-geometric quantities. The upshot is that the variation of
the $\l$-independent part of $X_0$ can be represented as a combined effect of
a coordinate variation of $\sK$ and a variation of $Y$.  Recall that the
automorphism transformations can be implemented by a coordinate
transformation (Remark \ref{coordinate} on page \pageref{coordinate}); then
$\sK$ and $Y$ should be brought to the new coordinate system by the
appropriate transformation. On top of that, there is a form-variation of $Y$,
equal simply to the infinitesimal function $\varphi$.

\smallskip

The use of the $\sK Y$ term can be illustrated in the Darboux coordinates; in
ref.~\cite{BM}, $X_0$ was given by \req{X00} satisfying \req{XmasterM} with
$R^{\a a}_0=0$.  Using in this case the Poisson bracket \req{Poissonbracket}
and differentiating  \req{38} with respect to $\pi^\a_a$ and $\Phi^\ast_{\a
a}$, we arrive at \BE
\dl{G_\b}{\Phi^\g}=-2\Bigl\{G_\b,\dl{Y}{\Phi^\g}\Bigr\}(-1)^{\e_\b\e_\g}\,,
\qquad \dl{G_\a}{\bar\Phi_\g}=
-2\Bigl\{G_\a,\dl{Y}{\bar\Phi_\g}\Bigr\}(-1)^{\e_\a\e_\g}\EE which constrains
the gauge functions in the $\Phi^\a,\,\bar\Phi_\a$-sector and is solved by
\BE Y=\bar\Phi_\a\Phi^\a - 2F(\Phi)\label{Yflat}\EE \BE
G_\a=\bar\Phi_\a-F(\Phi)\frac{\stackrel{\lea}{\d}}{\d\Phi^\a}
\label{Gflat}\EE where $F(\Phi)$ is a bosonic function of $\Phi^\a$ (`the
gauge {\it boson\/}').  This case illustrates clearly that the $G_\a$ must
not be annihilated by $V^a$. The gauge freedom contained in $N$ functions
$G_\a$ has eventually reduced to one bosonic function, which is needed for
gauge-fixing.

\pagebreak[3] \section{Concluding remarks\label{sec:7}}\lvm In this paper we
have developed the basics of antitriplectic differential geometry.  Based on
this, we have suggested a quantization of general hypergauge theories in an
$Sp(2)$-symmetric way in arbitrary coordinates on the field space.  This
generalizes what one has in the \asy\ case, which is also reviewed and
generalized here.  We have also seen that the most complete framework for the
hypergauge algebra generating relations is a triplectic extension of the
multilevel-type \cite{BT93-2,BT94-1} antibracket formalism.  It should be
realized that the formalism constructed  above is in fact developed in a
context much more general than the original motivation, which was the
$Sp(2)$-symmetric quantization and, in particular, the ghost-antighost
symmetry.

We have also given a reformulation of the usual antibracket scheme based on
the {\sl`weak'\/} master-equations, which resulted in allowing an a priori
arbitrary dependence of the gauge-fixing master-action on the `Lagrange
multipliers' $\l^\a$ (which do become true Lagrange multipliers to the
hypergauge conditions as soon as the action is taken to depend on them
linearly). This new possibility of introducing an arbitrary $\l$-dependence
is common to both the antisymplectic and triplectic cases, although in the
triplectic case the essence of the problem is stressed by the fact that
$\l$-independent terms exist in the master-action $X$ along with the
$\l$-dependent ones already at the classical level.  In the standard \asy\
quantization, on the other hand, $X_0\equiv X|_{\hbar=0}$ is homogeneous in
$\l^\a$, and only the `quantum' part (the measure $H$) is not proportional to
$\l^\a$. The difference between  \asy\ and triplectic cases is reflected in
the character of the Planck number conservation: in the triplectic case, only
a filtration by the Planck number is respected by the equations.

We have demonstrated that the partition function is independent of the
deformations of gauge-fixing master-action $X$ within a certain class allowed
by the algebra of hypergauge conditions.  The independence from the choice of
a solution of the master equation is what replaces gauge invariance in the
general case. The point is that in the general hypergauge theories the
gauge-fixing functions are hidden in the master-action $X$. This action is
`dual' to the master action $W$ whose expansion in antifields (once these are
identified) starts with the classical action that is gauge-independent in the
simplest sense of the word. On the other hand, the $X$ action is built
starting from {\it gauge-fixing\/} functions (and hence is tautologically
gauge-dependent).  It is quite remarkable that both these actions satisfy the
appropriate master equations. The freedom in choosing $X$ is then precisely
the independence from the gauge fixing.  The correct balance of degrees of
freedom is maintained when one considers the {\it maximal\/} arbitrariness in
solutions to the $X$-master-equation (such that is sufficient to establish
gauge independence in the usual sense).

We have been able to `lift' the formalism based on the weak master equation
to the extended triplectic manifold obtained by introducing triplectic
partners to the Lagrange multipliers $\l^\a$. This has allowed us to work
with the strong master equations and, in particular, to use the automorphism
formula. Moreover, strong master equations analogous to eq.~\req{semaster}
can be taken over to a tower of extended sectors in the antitriplectic
counterpart to the multilevel field-antifield formalism
\cite{BT93-1}--\cite{BT94-2}.  It will be shown elsewhere how the procedure
of reformulating the path integral on the extended triplectic manifold fits
into the general multilevel scheme obtained by gradually extending the phase
space by new variables ${\l^*}^{(n)}_{\a a}$, $\bar\l^{(n)}_\a$,
$\eta^{(n)}{}^{\a a}_{\phantom{Z}}$ and $\l^{(n+1)}{}^\a_{\phantom{Z}}$.

Several interesting questions have to do with the vector fields $V^a$ that we
have used to formulate the master equations.  One may e.$\,$g.\ ask if a
`reduction' (at least formal) is possible to the anticanonical case by simply
removing the superscripts ${}^a$ on $\Delta_\pm$ \req{Deltapm}, $V$ and the
antibrackets and dropping the terms with the $\epsilon$-symbols.  This would
result in the master equation of the form $\half(X,\,X) - V X= i\hbar\Delta
X$, with $V$ being an odd, nilpotent, operator that commutes with $\Delta$
and differentiates the antibracket. Such an equation was introduced recently
in \cite{SZ94} in a particular setting of string field theory, in the form of
the boundary operator (differential) $\d$.  It seems very interesting to
investigate the possibility of reformulating the general Lagrangian
quantization in the case when there is room for such a $\d$ to appear
(e.$\,$g\ when the antibracket degenerates).

There are several other interesting questions.  Recall that an action with
quadratic dependence on the `Lagrange multipliers' $\l$ is related to
non-singular gauges.  In the traditional gauge theories, there exists a
mechanism (a canonical transformation  \cite{BK}) allowing one to derive
these non-singular gauges from singular ones (imposed by delta-functions in
the integral).  It would be very interesting to understand higher orders in
$\l^\a$ in the triplectic case along these lines.  Another direction is a
possible triplectic generalization of the algebraic scheme
\cite{S,SchP,Zu,Ge} underlying \asy\ Lagrangian quantization.

\mbox{}

\pagebreak[3]

\subsection*{Acknowledgements} IAB and AMS are thankful to Lars Brink for
warm hospitality at the Institute of Theoretical Physics, Chalmers
University.  The work of IAB and AMS is partially supported by Human Capital
and Mobility program of the European Community under the Project
INTAS-93-2058.  AMS wishes to thank G.~Ferretti and I.~Tyutin for useful
comments.  IAB is also grateful to I.~Tyutin for stimulating discussions at
early stages of this work.  IAB was supported in part also by Board of
Trustees of ICAST and NATO Linkage Grant \#931717.

\newpage \appendix \section{Poisson bracket from the antibrackets}\lvm We are
going to show that, if one {\it defines\/} a $\{\;,\;\}$ by
\BE(F,\,V^aG)^b=\half\epsilon^{ab}\{F,\,G\}\,,\label{Poissondefinition}\EE
this would be a Poisson bracket on functions such that $(F,\,G)^a=0$ and
$(F,V^{\{a}G)^{b\}}=0$. This will mean that one can consistently relate the
LHS of \req{Poissondefinition} to the fixed Poisson bracket on the symplectic
submanifold $\cL_1$ introduced in section~\ref{sec:2.4}.

As mentioned in subsection~\ref{sec:2.4}, it follows from
\req{zeroantibracket} that \BE(F,V^aG)^b=-(-1)^{\e(F)\e(G)}(G,V^aF)^b\qquad
{\rm for}\quad F,G\in\cF(\cL_1)\,,\EE which gives the correct symmetry
properties.  Next, let us show that the derivation property holds:
\BE\new\BA{rl} \{F,GH\}&={}-\epsilon_{ab}(F,V^a(GH))^b\\
{}&={}-\epsilon_{ab}(F,(V^aG)H)^b - \epsilon_{ab}(-1)^{\e(G)}(F,G\,V^aH)^b\\
{}&={}-\epsilon_{ab}(F,V^aG)^b H -
\epsilon_{ab}(-1)^{\e(G)+\e(G)(\e(F)+1)}G\,(F,V^aH)^b\\
{}&={}\{F,G\}H+(-1)^{\e(F)\e(G)}G\{F,H\}\,.  \EA\EE

Now, we claim that the Jacobi identity for the Poisson bracket follows from
the condition \BE (F,(G,V^cH)^{\{b})^{d\}}=0\,\quad
F,G,H\in\cF(\cL_1)\label{three-condition}\EE which we have arrived at
in~\req{three}.  Applying $V^a$ to \req{three-condition}, we get
\BE\new\BA{rlr} \multicolumn{3}{l}{ (F,V^a(G,V^cH)^{\{b})^{d\}} }\\ \qquad
\qquad \qquad =&(-1)^{\e(F)}(V^aF,(G,V^cH)^{\{b})^{d\}}\\
=&(-1)^{\e(F)}((V^aF,G)^{\{b|},V^cH)^{|d\}}+
(-1)^{\e(F)\e(G)}(G,(V^aF,V^cH)^{\{b})^{d\}}&{\rm by\ \req{newJacobi}}\\
=&(-1)^{\e(H)\e(F)+\e(H)\e(G)+\e(F)\e(G)+\e(H)}
(V^cH,(G,V^aF)^{\{b})^{d\}}&{\rm by\ \req{symmetryproperty}}\\
{}&\multicolumn{2}{l}{ {}\!\!\!\!\new\BA{l} {}+(-1)^{\e(F)\e(G)}
(G,V^a(F,V^cH)^{\{b})^{d\}}\\ {}+(-1)^{\e(F)\e(G)+\e(F)}
(G,(F,V^aV^cH)^{\{b})^{d\}} \EA \hfill {\rm by\ \req{Vantibracket}} }\\
=&{}-\half(-1)^{\e(H)\e(G)+\e(H)\e(F)+\e(G)\e(F)}
(H,V^c\{G,F\})^{\{d}\epsilon^{b\}a}&{\rm by\ \req{three-condition}}\\
{}&{}-\half(-1)^{\e(F)\e(G)}(G,V^a\{F,H\})^{\{d}\epsilon^{b\}c}\\
{}&{}-\half(-1)^{\e(F)\e(G)+\e(F)}\epsilon^{ac}(G,(F,\sK H)^{\{b})^{d\}}&
{\rm by\ \req{cK}}\\
=&{}-\fourth(-1)^{\e(H)\e(G)+\e(H)\e(F)+\e(G)\e(F)}\{H,\{G,F\}\}
\epsilon^{c\{d}\epsilon^{b\}a}\\ {}&{}-\fourth(-1)^{\e(F)\e(G)}\{G,\{F,H\}\}
\epsilon^{a\{d}\epsilon^{b\}c}\\
{}&{}-\half(-1)^{\e(F)\e(G)+\e(F)}\epsilon^{ac}(G,(F,\sK H)^{\{b})^{d\}} \EA
\label{Jacobiderivation}\EE The LHS here, on the other hand, equals \BE
-\fourth\{F,\{G,H\}\}\epsilon^{a\{d}\epsilon^{b\}c}\EE Now the terms that
involve $\epsilon$-symbols and symmetrization in $bd$ are also symmetric in
$ac$, while the last term in \req{Jacobiderivation} is antisymmetric in $ac$
(and therefore has to vanish).  The equality among the $ac$-symmetric terms
is then equivalent to  the Jacobi identity for the Poisson bracket.

\section{Variations in the triplectic geometry}\lvm In contrast to their
(anti)symplectic counterparts, the triplectic antibrackets do vary under the
`canonical' transformations (which is largely due to the fact that the
`canonical' transformations are not truly canonical).  Consider an even
vector field $T\equiv T^A\d_A$ on our manifold $\cM$ (\asy\ or triplectic).
It will be convenient to define the Hamiltonian mapping from functions to
vector fields.  In the \asy\ case this is standard, \BE \oV:\,\cF(\cM)\to{\rm
Vect}(\cM):\, H\mapsto\oV_H\EE such that \BE \oV_H\,F=(H,F)\EE whence,
writing $\oV_H=\oV_H^A\d_A$, we find \BE
\oV_H^A=(-1)^{\e_A\e(H)+\e(H)+\e_A}E^{AB}\d_BH\EE where $E^{AB}$ determines
the antibracket by a formula directly analogous to \req{antiBG}.  In the
triplectic case, we have two such vector fields, \BE\new\BA{rcl}
\oV^a_H\,F&=&(H,F)^a\,,\\ \oV_H^{aA}&=&(-1)^{\e_A\e(H)+\e(H)+\e_A}E^{AB
a}\d_BH \EA\label{oVa}\EE For an odd $H$, $\oV^a_H$ are even, and it follows
that $\Delta^a H=-\half\div\oV^a_H$.  For $H$ even, $\oV_H$ is odd, and we
either define divergence as $\div V=\rho^{-1}\d_A(\rho V^A)$ or introduce the
components (as we did in the text) by writing $V=(-1)^{\e_A}V^A\d_A$; then,
$\Delta^a H=\half\div\oV^a_H$, and thus in any case \BE \Delta^a
H=\half(-1)^{\e(H)}\div\oV^a_H\,.\label{Deltadiv}\EE

Now, the variation of $\Delta^a$ under an infinitesimal change of variables
$\Gamma^A\mapsto\Gamma^A+T^A$ is given by the Lie derivative \BE
\sL_T\Delta^a=T\mycirc\Delta^a-\Delta^a\mycirc T \label{LieDelta}\EE and by
the same formula with the superscripts removed in the \asy\ case.  Using
\req{DeltaFG} to find the corresponding variation of the antibracket, we see
that $\sL_T(\,\cdot\,,\,\cdot\,)^a$ evaluates on two functions $F$, $G$ as
\BE T(F,\,G)^a-(TF,\,G)^a-(F,\,TG)^a\label{Tbracket}\EE which is nothing but
\BE \sL_T E^{ABa}=T^C\d_C E^{ABa} -T^A\!\stackrel{\lea}{\d}_C E^{CBa} -E^{AC
a}\d_C T^B\,.\label{LieE}\EE

Consider next the important case when the vector field $T$ is itself
hamiltonian.  In the \asy\ situation this means simply that $T=\oV_h$ for a
(fermionic) function $h$.  Then, $T$ differentiates the antibracket by virtue
of the Jacobi identity and therefore eq.~\req{Tbracket} (with the upper
indices dropped) vanishes.
In the triplectic case the situation is more interesting.  First, a
hamiltonian vector field $T=\oV_h$ is replaced by a sum
$T=\oV^1_{h_1}+\oV^2_{h_2}$.  In this case, the formula \req{LieDelta}
rewrites as \BE (\sL_T\Delta^a)(F)=(h_b,\,\Delta^aF)^b-\Delta^a(h_b,\,F)^b
\EE (recall that $h_a$ are fermionic). The
antibrackets, in turn, vary non-trivially due to the lack of Jacobi identity
relating $(\;,\;)^1$ and $(\;,\;)^2$:  $\sL_T(\;,\;)^a$ evaluates on two
functions $F$ and $G$ as
\BE(h_b,\,(F,\,G)^a)^b-((h_b,\,F)^b,\,G)^a-(F,\,(h_b,\,G)^b)^a\,.
\label{Lieantibracket}\EE

Next, the vector fields $V^a$ vary in the usual way:
\BE\sL_TV^a=[T,V^a]\,.\label{LieV}\EE Then, for a `hamiltonian' vector field
$T=\oV^a_{h_a}$ we find, using \req{Vantibracket},
\BE(\sL_TV^a)(F)=-(V^ah_c,\,F)^c\,.\EE

The formulae \req{LieV} and \req{Lieantibracket} have an interesting
consequence for the Poisson bracket defined as in \req{Poissondefinition}.
We find that the Lie derivative along $\oV^c_{h_c}$ of
$(\,\cdot\,,V^a\,\cdot\,)^b$ evaluates on a pair of functions $F$, $G$ on
$\cL_1$ as \BE\oV^c_{h_c}(F,\,V^aG)^b - (\oV^c_{h_c}F,\,V^bG)^a -
(F,\,V^b\oV^c_{h_c}G)^a\,.\label{B1}\EE Now we take here
$h_c=\epsilon_{cd}V^df$ with $f$ being a function on $\cL_1$, so that
$(F,\,f)^a=0$, $(F,\,V^af)^b=\half\epsilon^{ab}\{F,f\}$ and similarly for
$G$, which allows us to show that eq.~\req{B1} becomes \BE
\half\epsilon^{ab}\Bigl(-\{\{F,\,G\},\,f\}+
\{\{F,\,f\},\,G\}+\{F,\,\{G,\,f\}\} \Bigr)\EE which is the LHS of Jacobi
identity for the Poisson bracket and therefore vanishes.  The effect of the
chosen hamiltonian vector field is therefore that of a true hamiltonian
vector field on a symplectic manifold.  This suggests one of the possible
definitions of hamiltonian vector fields in the triplectic geometry as
$\epsilon_{ab}\oV^a_{V^b\varphi}$, which acts on functions as \BE
F\mapsto\epsilon_{ab}(V^b\varphi, F)^a\EE and is thus an extension to $\cM$
of the vector fields that are hamiltonian in the Poisson structure
on~$\cL_1$.

\bigskip

Consider next the form \req{X00} of the classical part of gauge-fixing
master-action. We are going to apply the above relations to derive the
transformation properties of the $\sK Y$ term from \req{X00}.  {}From
\req{LieV}, the variation of the operator $\sK$ \req{cK} generated by the
`hamiltonians' $h_a$ is given by
\BE(\sL_T\sK)(F)=-2\epsilon_{ab}(V^ah_c,\,V^bF)^c-(\sK h_c,\,F)^c
\label{LieK}\EE Now, once a master-action $X$ is chosen, there is a set of
hamiltonians $h_a$ depending on one arbitrary function as in
\req{hamiltonian}. The transformation generated by such $h_a$ is related by
\req{coordinatemapH} to the automorphism transformations of solutions to the
master equation. More precisely, since $\sK Y$ is the $\l$-independent part
of the master-action $X$, we pick out $\l$-independent terms in
\req{coordinatemapH}.  Thus we arrive at the hamiltonians \BE
h_c=-\epsilon_{cd}(\sK Y,\,\varphi)^d + 2\epsilon_{cd}V^d\varphi\EE
Remarkably, for $F=\sK Y$, the last term in \req{coordinatemapH} drops out
due to \req{projection}, and we are left with a purely `hamiltonian' piece!

Now we can evaluate \req{LieK} with the above $h_a$: for
$\delta\sK\equiv\sL_T\sK$, we find \BE (\delta\sK)(F)= -\epsilon_{ab}((\sK
Y,\,\varphi)^b,\,\sK F)^a -2\epsilon_{ab}\sK(\varphi,\,V^bF)^a
+2\epsilon_{ab}V^b(\varphi,\,\sK F)^a + \epsilon_{ab}\sK((\sK
Y,\,\varphi)^b,\,F)^a\label{formFold}\EE where we have used
\req{Vantibracket} and the projector property \req{projection} to pull some
of the $V^a$ out of the antibrackets.

Returning now to \req{X00}, we extract from the automorphism transformations
from section~\ref{sec:5} the variation of $\sK Y$.  As in \req{241},
\req{242}, we will consider for simplicity only the `main' part of the
variation, given by the antibracket and $V^a$. Thus, the automorphism formula
tells us that the $\lambda$-independent terms in \req{X00} vary as \BE
\delta(\sK Y) = -\epsilon_{ab}((\sK Y,\,\varphi)^b,\,\sK Y)^a -
2\epsilon_{ab}V^a(\varphi,\,\sK Y)^b + \sK\varphi\EE Since the automorphism
can be implemented by a coordinate transformation, and as long as the
coordinate transformations do not leave $\sK$ invariant, we have to transform
$\sK$ to the new coordinate system.  Subtracting from the last expression the
effect on $\sK$ of the coordinate transformation \req{formFold} with $F=Y$,
we find \BE \delta(\sK Y) - (\delta\sK)(Y)= \sK\varphi +
2\epsilon_{ab}\sK(\varphi,\,V^bY)^a -\epsilon_{ab}\sK((\sK
Y,\,\varphi)^b,\,Y)^a\EE which {\it is\/} of the desired form $\sK(\ldots)$.
Moreover, the variation of $Y$ that follows from this formula can be written
as \BE \delta Y =\varphi +\epsilon_{ab}(Y,\,(\sK Y,\,\varphi)^b)^a +
2\epsilon_{ab}(Y,\,V^a\varphi)^b\EE where the two last terms are, again, the
effect of the hamiltonian transformation generated by \req{hamiltonian}, that
is, the $\l$-independent part of the coordinate transformation
\req{coordinatemap} (we have used the projector property \req{projection}).
The remaining term $\varphi$  is a {\it form\/}variation of $Y$. This
establishes the `covariance' properties and hence consistency of the ansatz
\req{X00} for the master-action $X$ that leads to singular (delta-function)
gauges.

\def\JMP{{\sl J.\ Math.\ Phys.}}
\def\NPB{{\sl Nucl. Phys}. B}
\def\PLB{{\sl Phys. Lett}. B}
\def\MPLA{{\sl Mod. Phys. Lett}. A}
\def\CMP{{\sl Commun. Math. Phys.}}
\def\IJMPA{{\sl Int. J. Mod. Phys}. A}
\def\MPLA{{\sl Mod. Phys. Lett}. A}
\def\PRD{{\sl Phys. Rev}. D}

\pagebreak[4]

\end{document}